\documentclass[10pt,a4paper]{article}

\usepackage{jheparxiv}
\usepackage[T1]{fontenc}
\usepackage{tikz}
\usepackage{tikz-cd}
\usepackage{ctable}

\DeclareMathOperator{\Tr}{Tr}

\newcommand{\cO}{\mathcal{O}}

\newcommand{\rd}{\mathrm{d}}
\newcommand{\p}{\partial}

\def\tb{b_{0}{}}
\def\tg{g_{0}{}}
\def\th{h_{0}{}}
\def\ts{s_{0}{}}
\def\tv{\tilde{v}}
\def\tA{A_{0}{}}
\def\tB{B_{0}{}}

\def\tE{E_{0}{}}
\def\tV{\tilde{V}}

\def\hg{\hat{g}}
\def\hl{\hat{l}}
\def\hbarl{\hat{\bar{l}}}

\def\hs{\hat{s}}

\def\hB{\hat{B}}
\def\hE{\hat{E}}
\def\hM{\hat{M}}
\def\hN{\hat{N}}

\def\hmu{\hat{\mu}}
\def\hnu{\hat{\nu}}
\def\hrho{\hat{\rho}}
\def\hsigma{\hat{\sigma}}
\def\hphi{\hat{\phi}}

\def\htg{\hat{g}_{0}{}}
\def\hts{\hat{s}_{0}{}}

\def\hcH{\hat{\mathcal{H}}}
\def\hcJ{\hat{\mathcal{J}}}
\def\hcO{\hat{\mathcal{O}}}

\def\cA{\mathcal{A}}
\def\barcA{\bar{\mathcal{A}}}

\newcommand{\nn}{\nonumber}

\title{ The Classical Double Copy for Half-Maximal Supergravities and T-duality}

\author[a]{Stephen Angus}
\author[a]{\!\!, Kyoungho Cho}
\author[a,b]{and Kanghoon Lee}

\affiliation[a]{Asia Pacific Center for Theoretical Physics, Postech, Pohang 37673, Korea}
\affiliation[b]{Department of Physics, Postech, Pohang 37673, Korea}

\emailAdd{stephen.angus@apctp.org}
\emailAdd{kyoungho.cho@apctp.org}
\emailAdd{kanghoon.lee1@gmail.com}

\abstract{We study the classical double copy for ungauged half-maximal supergravities using the Kaluza--Klein reduction of double field theory (DFT). We construct a general formula for the Kaluza--Klein (KK) reduction of the DFT Kerr--Schild ansatz. The KK reduction of the ansatz is highly nonlinear, but the associated equations of motion are linear. This linear structure implies that half-maximal supergravities admit a classical double copy. We show that their single copy is given by a pair of Maxwell-scalar theories, which are the KK reduction of a higher-dimensional single copy of DFT. We also investigate their T-duality transformations --- both the Buscher rule and continuous $\mathit{O}(D,D)$ rotations. Applying the Buscher rule to the Kerr BH, we obtain a solution with a nontrivial Kalb--Ramond field and dilaton. We also identify the single copy of Sen's heterotic BH and the chiral null model and show that the chiral null model is self-dual under T-duality rotations.}

\begin{document}
\begin{flushright}
APCTP Pre2021 - 010
\end{flushright}

\maketitle

\section{Introduction}
As a theory of quantum gravity, string theory exhibits an intriguing relation between gravity and gauge theory. This relation is known as the double copy \cite{Kawai:1985xq,Bern:2008qj,Bern:2010ue,BjerrumBohr:2009rd,Stieberger:2009hq,Bern:2010yg,BjerrumBohr:2010zs,Feng:2010my,Tye:2010dd,Mafra:2011kj,Monteiro:2011pc,BjerrumBohr:2012mg} and is considered an important guiding principle in quantum gravity. The closed string mode expansion is divided into left and right movers, and each mode is associated with an open string. From the perspective of scattering amplitudes, closed string (gravity) amplitudes are represented by squaring amplitudes for open strings (gauge theory). This is known as the KLT relation \cite{Kawai:1985xq} and is equivalent to the double copy for tree-level scattering amplitudes.

In recent years there has been substantial progress in extending the double copy prescription to exact solutions of the classical equations of motion (perturbative classical solutions were explored in \cite{Anastasiou:2014qba,Borsten:2015pla,Anastasiou:2016csv,Goldberger:2016iau,Luna:2016hge,Goldberger:2017frp,Anastasiou:2017nsz,Cardoso:2016ngt,Borsten:2017jpt,Anastasiou:2017taf,Anastasiou:2018rdx,LopesCardoso:2018xes,Shen:2018ebu,Carrillo-Gonzalez:2018pjk,Plefka:2018dpa}).  This is the so-called \textit{classical double copy} \cite{Monteiro:2014cda,Luna:2015paa,Luna:2016due,Lee:2018gxc,Berman:2018hwd,Gurses:2018ckx,Luna:2018dpt}, which provides an explicit map between gravity and gauge fields through the Kerr--Schild (KS) ansatz. The KS ansatz has been useful in finding physically important solutions to general relativity. The power of the ansatz lies in the fact that it linearises Einstein's equations and reduces the resulting linear equations to Maxwell's equations, paving the way for a simple description of complex gravitational solutions in terms of well-understood electromagnetism. Other relevant recent work is in \cite{Sabharwal:2019ngs,Alawadhi:2019urr,Kim:2019jwm,Banerjee:2019saj,Bahjat-Abbas:2020cyb,Alfonsi:2020lub,Keeler:2020rcv,Elor:2020nqe,Momeni:2020vvr,Alawadhi:2020jrv,Godazgar:2020zbv,Prabhu:2020avf,Carrasco:2020ywq,Ferrero:2020vww,Chacon:2020fmr,White:2020sfn,Monteiro:2020plf,Lescano:2021ooe,Alkac:2021bav,Lescano:2021but,Campiglia:2021srh,Chacon:2021wbr,Farnsworth:2021wvs,Alkac:2021seh}.

Recently the KS ansatz and the classical double copy in GR have been extended to NSNS gravity through the use of double field theory (DFT). DFT is a low energy effective field theory of closed string theory with manifest $\mathit{O}(D,D)$ T-duality \cite{Siegel:1993bj,Siegel:1993xq,Hull:2009mi,Hohm:2010jy,Hohm:2010pp}. It has been shown that DFT and the double copy share the same features originating from string theory \cite{Hohm:2011dz,Cheung:2016say,Cheung:2017kzx,Lee:2018gxc,Cho:2019ype,Kim:2019jwm}, in particular, the factorisation into left- and right-moving sectors.  The correspondence has recently been extended to M-theory via exceptional field theory, which is an M-theory extension of DFT \cite{Berman:2020xvs}. Thus DFT provides a natural framework for describing the double copy.

The double copy has been extended to a larger class depending on the number of supersymmetries and multiplets --- including maximal and half-maximal SUSY \cite{Bern:2008qj,Bern:2010ue,Bern:2010yg,Bern:2011rj,Bern:2009kd,Bern:2013uka,Bern:2014sna} --- from the scattering amplitude point of view. On the other hand, the classical double copy for half-maximal supergravities is thus far not yet established due to the absence of the relevant KS formalism and the corresponding linear structure of the KS equations of motion. One aim of this paper is to construct the classical double copy for ungauged half-maximal supergravities. Note that Kaluza--Klein (KK) reduction or toroidal compactification of ten-dimensional $N=1$ supergravities yields ungauged half-maximal supergravities in lower dimensions. Here we will consider the minimal ten-dimensional $N=1$ supergravity without any additional vector multiplets. Then the ten-dimensional theory can be embedded into $\mathit{O}(D,D)$ DFT, where $D$ is the dimensionality of the theory, but for generality we will not fix the value of $D$.

Since the KK reduction of a higher-dimensional DFT defines ungauged half-maximal supergravities, the lower-dimensional KS ansatz can be constructed from the higher-dimensional one. We will show that the KK reduction of the higher-dimensional KS ansatz is highly nonlinear and messy; however, the equations of motion remain linear. This linear structure enables us to spell out the classical double copy after KK reduction. Thus the single copy of the lower-dimensional half-maximal supergravities is given by the KK reduction of a pair of higher-dimensional Maxwell theories: Maxwell-scalar theories.  As in higher-dimensional DFT, the left- and right-mover decomposition plays a crucial role in the double copy.

Another interesting aspect of the ungauged half-maximal supergravity is T-duality. T-duality arises from string theory on a toroidal background and is a powerful solution-generating technique in supergravity. Many important solutions can be generated via a duality transformation from a simpler seed solution. We will discuss not only the Buscher rule \cite{Buscher:1987qj,Rocek:1991ps}, but also continuous $\mathit{O}(D,D)$ rotations. We will present the T-duality transformation of the classical double copy for those solutions. In doing so, we will see that DFT plays a crucial role because of its manifest T-duality covariance. In the four-dimensional case with a timelike Killing vector, the duality transform is realised via mixing of the electric and scalar charges.

We will study three examples. First of all, we will consider the Buscher transformation of the Kerr black hole (BH). This  provides a solution without horizon involving a nontrivial $B$-field and dilaton. The single copy of this solution is given by the single copy of the Kerr BH, in which the sign of one of the electric charges is flipped. Following that, we will identify the single copy for Sen's heterotic BH \cite{Sen:1994eb}, which is given by a T-duality rotation of the Kerr BH. Finally, we will show that the KK-reduced chiral null model \cite{Horowitz:1994rf,Behrndt:1994tf,Behrndt:1995tr,Tseytlin:1995fh} is self-dual under $\mathit{O}(n,n)$ transformations and identify its single copy.

This paper is organised as follows. In section \ref{sec:2} we obtain the KK reduction of the KS ansatz. We show that the equations of motion are linear even after the KK reduction. In section \ref{sec:3} we discuss the T-duality transformation of the KS ansatz, including both the Buscher rule and the continuous $\mathit{O}(n,n)$ transformation. Section \ref{sec:4} describes the single and double copy for half-maximal supergravities using the KS ansatz. In section \ref{sec:5} we study the single copy for three examples as an application: the Buscher rule for the Kerr BH, Sen's rotating BH in four-dimensional heterotic supergravity, and the chiral null model. Section \ref{sec:6} gives our conclusions.

\section{Kaluza--Klein reduction of the Kerr--Schild ansatz}\label{sec:2}

Ungauged half-maximal supergravities can be constructed via the Kaluza--Klein reduction of ten-dimensional NSNS supergravity (with or without vector multiplets). If there are $n$ additional vector multiplets, then the ten-dimensional supergravity can be embedded into $\mathit{O}(10,10+n)$ heterotic DFT. Throughout this paper, we will ignore vector multiplets for simplicity.  Since the structure of the bosonic sector of DFT is independent of the spacetime dimension, for generality we will consider  $\mathit{O}(D,D)$ DFT. First, we review the Kaluza--Klein (KK) reduction of DFT and NSNS supergravity and fix our notation. We then investigate the KK reduction of the Kerr--Schild (KS) ansatz of DFT. Finally, we discuss the KK reduction of the KS equations of motion and show that these remain linear, even though the lower-dimensional KS ansatz is nonlinear even at the level of DFT.

\subsection{Kaluza--Klein reduction of DFT} \label{sec:2.1}

Let us consider a $D$-dimensional spacetime $M_{D}$ endowed the massless NSNS sector fields, namely, the metric $\hat{g}$, 2-form gauge field $\hat{B}$ and scalar $\hat{\phi}$.\footnote{Throughout this article, all hatted objects stand for quantities defined in the $D$-dimensional 
total space (or its double).} Assume that $M_{D}$ is a torus bundle over a $d$-dimensional base space $N_{d}$ with a fibre $T^{n}$, where $D=d+n$, and introduce coordinates $z^{\hat{\mu}}$ on a local patch of $M_{D}$.
We require that the $T^{n}$ fibre allows $n$ abelian Killing vectors $k_{\alpha}{}^{\hat{\mu}}$ for KK reduction.   Then introduce a local bundle patch $\hat{U}$ for $M_{D}$, which is constructed from a direct product $\hat{U} = U \times T^{d}$, where $U$ is a patch on the base $N_{d}$, and define coordinates $\hat{x}^{\hat{\mu}}$ on $\hat{U}$ which decompose into the base coordinates $x^{\mu}$ and the $T^{n}$ fibre coordinates $y^{a}$,
\begin{equation}
  \hat{x}^{\hat{\mu}} = (x^{\mu} \,, y^{\alpha})\,.
\label{bundle_coordinates}\end{equation}
Note that the $z^{\hat{\mu}}$ do not have to be the same as the $\hat{x}^{\hat{\mu}}$, but we can always find a diffeomorphism identifying them. We may set the coordinates on $T^{n}$ to be those in which the Killing vectors become constant, $k_{\alpha}{}^{\hat{\mu}} = \delta_{\alpha}{}^{\hat{\mu}}$, such that
\begin{equation}
  k_{\alpha} = \frac{\partial}{\partial y^{\alpha}} \, .
\label{}\end{equation}
The associated Lie derivative is the partial derivative with respect to $y^{\alpha}$, and all fields are independent of $y^{\alpha}$. This procedure thus naturally accommodates the KK reduction from $M_{D}$ to $N_{d}$. Here we treat $N_{d}$ as an external space and $T^{n}$ as an internal space after the KK reduction.

The torus metric is given by $h_{\alpha\beta} = k_{\alpha}{}^{\hat{\mu}} k_{\beta}{}^{\hat{\nu}} \hat{g}_{\hat{\mu}\hat{\nu}} = \hat{g}_{\alpha\beta}$. The one form fields $k^{\alpha} = h^{\alpha\beta} k_{\beta}$ are represented by
\begin{equation}
  k^{\alpha} = \mathrm{d}y^{\alpha} + A_{\mu}{}^{\alpha}\mathrm{d}x^{\mu} \, ,
\label{}\end{equation}
where $A^{\alpha}$ are the connection of the torus bundle $M_{D}$. The KK ansatz for the massless NSNS sector is given by
\begin{equation}
\begin{aligned}
  \hat{g}_{\hat{\mu} \hat{\nu}} &= \begin{pmatrix}
     g_{\mu\nu} + A_{\mu}{}^{\alpha} h_{\alpha\beta} A^{\beta}{}_{\nu} & A_\mu{}^{\gamma} h_{\gamma\beta} \\
     h_{\alpha\gamma} A^{\gamma}{}_{\nu} & h_{\alpha\beta}
   \end{pmatrix} \,,
  \\
  \hat{B}_{\hat{\mu} \hat{\nu}} &= \begin{pmatrix}
     B_{\mu\nu} + \frac{1}{2} \big( C_{\mu\alpha}(A^{t})^{\alpha}{}_{\nu}- A_{\mu}{}^{\alpha} (C^{t})_{\alpha\nu}\big) + A_{\mu}{}^{\alpha} b_{\alpha\beta} (A^{t})^{\beta}{}_{\nu}  &~ C_{\mu\beta} + A_{\mu}{}^{\alpha} b_{\alpha\beta} \\
    -(C^{t})_{\alpha\nu} + b_{\alpha\beta} (A^{t})^{\beta}{}_{\nu} & b_{\alpha\beta}
  \end{pmatrix}\,,
  \\
  \hat{\phi} &= \phi +\frac{1}{4} \ln \det h \,.
\end{aligned}\label{KKansatz_gBphi}
\end{equation}
After the KK reduction, $A^{\alpha}$ and $C_{\alpha}$ become $\mathit{U}(1)$ gauge fields, and $h$ and $b$ become scalar fields.

To understand the KK reduction in the context of DFT, we embed the torus bundle into a $2D$-dimensional doubled space $\mathcal{M}_{2D}$. We introduce a $2D$-dimensional total space consisting of a $2d$-dimensional \emph{doubled external space} $\mathcal{N}_{2d}$ and a complementary $2n$-dimensional \emph{internal doubled torus} $T^{2n}$. We denote the doubled coordinates of $\mathcal{M}_{2D}$ as
\begin{equation}
  \hat{X}^{\hat{M}} = \{\hat{x}^{\hat{\mu}}\,,\hat{\tilde{x}}_{\mu}\}\,, \qquad \hat{M},\hat{N} \cdots = 1,2,\cdots, 2D \, ,
\end{equation}
where $\hat{x}^{\hat{\mu}}$ are the coordinates of $M_D$ and $\tilde{x}_{\hat{\mu}}$ are their dual coordinates. An $\mathit{O}(D,D)$ vector $V_{\hat{M}}$ unifies a $D$-dimensional vector $\hat{v}^{\hat{\mu}}$ and a one-form $\hat{k}_{\hat{\mu}}$ in an $\mathit{O}(D,D)$ covariant way,
\begin{equation}
  \hat{V}_{\hat{M}} = \begin{pmatrix} \hat{v}^{\hat{\mu}} \\ \hat{k}_{\hat{\mu}} \end{pmatrix} \,,
  \qquad
  \hat{V}^{\hat{M}} = \begin{pmatrix} \hat{k}_{\hat{\mu}} \\ \hat{v}^{\hat{\mu}} \end{pmatrix} \,.
\label{}\end{equation}
For consistency of the theory, we need to impose the section condition,
\begin{equation}\label{key}
  \Big(\partial_{\hat{M}} \bullet\Big) \Big( \partial^{\hat{M}} \bullet\Big) = 0\,,
\end{equation}
which we satisfy by imposing independence of $\hat{\tilde{x}}_{\hat{\mu}}$, i.e. $\frac{\partial}{\partial \hat{\tilde{x}}_{\hat{\mu}}} = \tilde{\partial}^{\hat{\mu}}=0$. The total space coordinates $\hat{X}^{\hat{M}}$ are decomposed into the external part $X^{M}$, where $M,N,\cdots=1,2,\cdots 2d$, and the internal part $Y^{A}$, where $A,B,\cdots 1,2,\cdots, 2n$
\begin{equation}
  \hat{X}^{\hat{M}} = \{X^{M}\,, Y^{A}\}
\label{}\end{equation}
As before, we assume that all fields are independent of the internal coordinates $Y^{A}$, $\partial_{A} = 0$.

The field content of $\mathit{O}(D,D)$ DFT defined in $\mathcal{M}_{2D}$ consists of the generalised metric $\hat{\mathcal{H}}_{\hat{M}\hat{N}}$ and the DFT dilaton $\hat{d}$. The generalised metric is a symmetric $\mathit{O}(D,D)$ element satisfying the so-called $\mathit{O}(D,D)$ constraint,
\begin{equation}
  \hat{\mathcal{H}}_{\hat{M}\hat{N}} \hat{\mathcal{J}}^{\hat{N}\hat{P}} \hat{\mathcal{H}}_{\hat{P}\hat{Q}} = \hat{\mathcal{J}}_{\hat{M}\hat{Q}}\,,
\label{ODDconstraint}\end{equation}
where $\hat{\mathcal{J}}$ is the $\mathit{O}(D,D)$ metric parametrised by
\begin{equation}
  \hat{\mathcal{J}}_{\hat{M}\hat{N}} = \begin{pmatrix} 0 & \delta^{\hat{\mu}}{}_{\hat{\nu}} \\ \delta_{\hat{\mu}}{}^{\hat{\nu}} & 0\end{pmatrix}\,.
\label{}\end{equation}
Note that $\hat{J}$ defines the inner product in the generalised tangent space, and thus also raises and lowers the $\mathit{O}(D,D)$ vector indices, $\hat{V}^{\hat{M}} = \hat{J}^{\hat{M}\hat{N}} \hat{V}_{\hat{N}}$ and $\hat{V}_{\hat{M}} = \hat{J}_{\hat{M}\hat{N}} \hat{V}^{\hat{N}}$.
We may solve the $\mathit{O}(D,D)$ constraint \eqref{ODDconstraint} explicitly and find a parametrisation of $\hat{\mathcal{H}}$ in terms of $\hat{g}$ and $\hat{B}$, given by
\begin{equation}
  \hat{\mathcal{H}}_{\hat{M}\hat{N}} = \begin{pmatrix} \hat{g}^{\hat{\mu}\hat{\nu}} & -\hat{g}^{\hat{\mu}\hat{\rho}} \hat{B}_{\hat{\rho}\hat{\nu}} \\ \hat{B}_{\hat{\mu}\hat{\rho}} \hat{g}^{\hat{\rho}\hat{\nu}} &~ \hat{g}_{\hat{\mu}\hat{\nu}} - \hat{B}_{\hat{\mu}\hat{\rho}} \hat{g}^{\hat{\rho}\hat{\sigma}} \hat{B}_{\hat{\sigma}\hat{\nu}} \end{pmatrix}\,.
\label{para_hatH}\end{equation}

Let us consider the KK reduction of $\mathit{O}(D,D)$ DFT to $\mathcal{N}_{2d}$ by compactifying on the internal space $T^{2n}$. The KK ansatz for the generalised metric is given by \cite{Berman:2013cli}
\begin{equation}
\hat{\mathcal{H}}_{\hat{M} \hat{N}} = \begin{pmatrix}
  \hat{\mathcal{H}}_{MN} & \hat{\mathcal{H}}_{M B} \\
  \hat{\mathcal{H}}_{AN} & \hat{\mathcal{H}}_{A B}
\end{pmatrix}\,,
\label{KK_gen_metric1}\end{equation}
where
\begin{equation}
\begin{aligned}
  \hat{\mathcal{H}}_{MN} &= \Big(\delta_{M}{}^{P} -\frac{1}{2} W_{M}{}^{A} (W^t)_{A}{}^{P} \Big) \mathcal{H}_{PQ} \Big(\delta^{Q}{}_{N} -\frac{1}{2} W^{Q}{}_{B} (W^t)^{B}{}_{N} \Big) + W_M{}^{A} \mathcal{M}_{A B} (W^{t})^{B}{}_{N} \, ,
  \\
  \hat{\mathcal{H}}_{M B} &= \Big(\delta_{M}{}^{P} -\frac{1}{2} W_{M}{}^{A} (W^t)_{A}{}^{P} \Big)  \mathcal{H}_{PN} W^{N}{}_{B} - W_M{}^{A} \mathcal{M}_{A B} \, ,
  \\
  \hat{\mathcal{H}}_{AB} &= \mathcal{M}_{A B} + (W^t)_{A}{}^{M} \mathcal{H}_{MN} W^{N}{}_{B}\,,
\end{aligned}\label{KK_gen_metric2}
\end{equation}
where $\mathcal{H}_{MN}$ and $\mathcal{M}_{AB}$ are the $\mathit{O}(d,d)$ and $\mathit{O}(n,n)$ generalised metrics, respectively. As in \eqref{para_hatH}, these are parametrised in terms of the supergravity fields in the external space $\mathcal{N}_{2d}$ and the internal space $T^{2n}$ as
\begin{equation}
\begin{aligned}
  \mathcal{H}_{MN} = \begin{pmatrix} g^{\mu\nu} & -g^{\mu\rho} B_{\rho\nu} \\ B_{\mu\rho} g^{\rho\nu} & g_{\mu\nu} - B_{\mu\rho} g^{\rho\sigma} B_{\sigma\nu} \end{pmatrix}\,,
  \qquad
  \mathcal{M}_{AB} = \begin{pmatrix} h^{\alpha\beta} & -h^{\alpha\gamma} b_{\gamma\beta} \\ b_{\alpha\gamma} h^{\gamma\beta} & h_{\alpha\beta} - b_{\alpha\gamma} h^{\gamma\delta} b_{\delta\beta}\end{pmatrix}\,,
\end{aligned}\label{}
\end{equation}
and $W_{MA}$ is the set of doubled $\mathit{U}(1)$ gauge fields that are associated with the off-diagonal components of $\hat{g}$ and $\hat{B}$ in \eqref{KKansatz_gBphi},
\begin{equation}
  W_{M A} = \begin{pmatrix} 0 & 0\\ A_{\mu}{}^{\alpha} & C_{\mu \alpha}\end{pmatrix}\,.
\label{para_W}\end{equation}

The characteristic feature of DFT that connects it to the double copy is the double local Lorentz group
\begin{equation}
  \mathit{O}(1,D-1)_{L}\times\mathit{O}(1,D-1)_{R} \,,
\label{double_local_Lorentz}\end{equation}
which is a Lorentizian version of the maximal compact subgroup of $\mathit{O}(D,D)$. It arises from the left-right mover decomposition of the closed-string mode expansion and shares a common origin with the KLT relation \cite{Kawai:1985xq}. This structure leads to the double-vielbein or generalised frame fields denoted as $\hat{V}_{\hat{M}}{}^{\hat{m}}$ and $\hat{\bar{V}}_{\hat{M}}{}^{\hat{\bar{m}}}$, where $\hat{m}$ and $\hat{\bar{m}}$ are $\mathit{O}(1,D-1)_{L}$ and $\mathit{O}(1,D-1)_{R}$ local frame indices, respectively. These satisfy the defining conditions
\begin{equation}
  \hat{V}_{\hat{M}\hat{m}} \eta^{\hat{m}\hat{n}}(\hat{V}^{t})_{\hat{n}\hat{N}}= \hat{P}_{\hat{M} \hat{N}}\,, \qquad \hat{\bar{V}}_{\hat{M} \hat{\bar{m}}} \bar{\eta}^{\hat{\bar{m}}\hat{\bar{n}}}(\hat{\bar{V}}^{t})_{\hat{\bar{n}}\hat{N}} =- \hat{\bar{P}}_{\hat{M} \hat{N}}\,,
\label{defining_VVbar}\end{equation}
where $\hat{P} = \frac{1}{2}(\hat{\mathcal{J}} + \hat{\mathcal{H}})$ and $\hat{\bar{P}} = \frac{1}{2}(\hat{\mathcal{J}} - \hat{\mathcal{H}})$.
They are parametrised in terms of the supergravity fields by solving the above conditions, yielding
\begin{equation}
  \hat{V}_{\hat{M}}{}^{\hat{m}} = \frac{1}{\sqrt{2}} \begin{pmatrix} \hat{e}^{\hat{\mu} \hat{m}}\\ (\hat{g}+\hat{B})_{\hat{\mu}\hat{\nu}}\, \hat{e}^{\hat{\mu}\hat{m}} \end{pmatrix}\,,
  \qquad
  \hat{\bar{V}}_{\hat{M}}{}^{\hat{\bar{m}}} = \frac{1}{\sqrt{2}} \begin{pmatrix} \hat{\bar{e}}^{\hat{\mu} \hat{\bar{m}}}\\ (-\hat{g}+\hat{B})_{\hat{\mu}\hat{\nu}}\, \hat{\bar{e}}^{\hat{\mu}\hat{\bar{m}}} \end{pmatrix}\,,
\label{}\end{equation}
where $\hat{e}$ and $\hat{\bar{e}}$ are the usual background vielbeins associated with the common metric $\hat{g}$,
\begin{equation}
  \hat{e}_{\hat{\mu}}{}^{\hat{m}} \hat{e}_{\hat{m} \hat{\nu}} = \hat{\bar{e}}_{\hat{\mu}}{}^{\hat{\bar{m}}} \hat{\bar{e}}_{\hat{\bar{m}} \hat{\nu}}=\hat{g}_{\hat{\mu} \hat{\nu}} \,.
\label{}\end{equation}

For the KK reduction of the double-vielbein, we decompose the generalised frame indices into the external part, $\mathit{O}(1,d-1)_{L}\times \mathit{O}(1,d-1)_{R}$, and the internal part, $\mathit{O}(n)\times \mathit{O}(n)$, as
\begin{equation}
  \hat{m} = \{m,a\}\,, \qquad \hat{\bar{m}} = \{\bar{m},\bar{a}\}\,.
\label{}\end{equation}
The KK ansatzes for the usual vielbeins, $\hat{e}$ and $\hat{\bar{e}}$, are
\begin{equation}
  \hat{e}_{\hat{\mu}}{}^{\hat{m}}=\begin{pmatrix} e_{\mu}{}^{m} & A_{\mu}{}^{\alpha} \Phi_{\alpha}{}^{a} \\ 0 & \Phi_{\alpha}{}^{a} \end{pmatrix} \,,
  \qquad
  \hat{\bar{e}}_{\hat{\mu}}{}^{\hat{\bar{m}}}=\begin{pmatrix} \bar{e}_{\mu}{}^{\bar{m}} & A_{\mu}{}^{\alpha} \bar{\Phi}_{\alpha}{}^{\bar{a}} \\ 0 & \bar{\Phi}_{\alpha}{}^{\bar{a}} \end{pmatrix} \,,
\label{KK_ansatz_vielbein}\end{equation}
where
\begin{equation}
  e_{\mu}{}^{m} e_{m\nu} = \bar{e}_{\mu}{}^{\bar{m}} \bar{e}_{\bar{m}\nu} = g_{\mu\nu}\,,
  \qquad
  \Phi_{\alpha}{}^{a} \Phi_{a\beta} = \bar{\Phi}_{\alpha}{}^{\bar{a}} \bar{\Phi}_{\bar{a}\beta} = h_{\alpha\beta}\,.
\label{}\end{equation}
One can easily check that \eqref{KK_ansatz_vielbein} is consistent with the KK ansatz of the metric \eqref{KKansatz_gBphi}. Similarly, the KK reduction ansatz for the double-vielbein, which is consistent with the KK reduction of the generalised metric \eqref{KK_gen_metric1} and \eqref{KK_gen_metric2}, is given by \cite{Berman:2013cli}
\begin{equation}
\begin{aligned}
  \hat{V}_{\hat{M}}{}^{\hat{m}} &= \begin{pmatrix} V_{M}{}^{m}-\frac{1}{2} W_{M}{}^{A} (W^{t})_{A}{}^{N} V_{N}{}^{m} &~ -W_{M}{}^{A} V_{A}{}^{a} \\  (W^{t})_{A}{}^{M} V_{M}{}^{m} &V_{A}{}^{a}\end{pmatrix} \,,
  \\
  \hat{\bar{V}}_{\hat{M}}{}^{\hat{\bar{m}}} &= \begin{pmatrix} \bar{V}_{M}{}^{\bar{m}}-\frac{1}{2} W_{M}{}^{A} (W^{t})_{A}{}^{N} \bar{V}_{N}{}^{\bar{m}} & ~ -W_{M}{}^{A} \bar{V}_{A}{}^{\bar{a}} \\  (W^{t})_{A}{}^{M} \bar{V}_{M}{}^{\bar{m}} & \bar{V}_{A}{}^{\bar{a}} \end{pmatrix} \,,
\end{aligned}\label{KK_DV}
\end{equation}
where $V_{M}{}^{m}$ and $\bar{V}_{M}{}^{\bar{m}}$ are $\mathit{O}(d,d)$ double-vielbeins and $V_{A}{}^{a}$ and $\bar{V}_{A}{}^{\bar{a}}$ are $\mathit{O}(n,n)$ double-vielbeins,
\begin{equation}
\begin{aligned}
  V_{M}{}^{m} &= \frac{1}{\sqrt{2}} \begin{pmatrix} e^{\mu m} \\ (g+B)_{\mu\nu}e^{\nu m}\end{pmatrix}\,,
  &\qquad
  \bar{V}_{M}{}^{\bar{m}} &= \frac{1}{\sqrt{2}} \begin{pmatrix} \bar{e}^{\mu\bar{m}} \\ (-g+B)_{\mu\nu}e^{\nu\bar{m}}\end{pmatrix}\,,
  \\
  V_{A}{}^{a} &= \frac{1}{\sqrt{2}} \begin{pmatrix} \Phi^{\alpha a} \\ (h+b)_{\alpha\beta}\Phi^{\beta a}\end{pmatrix}\,,
  &\qquad
  \bar{V}_{A}{}^{\bar{a}} &= \begin{pmatrix} \bar{\Phi}^{\alpha \bar{a}} \\ (-h+b)_{\alpha\beta}\bar{\Phi}^{\beta \bar{a}}\end{pmatrix}\,.
\end{aligned}\label{}
\end{equation}
%

\subsection{Kaluza--Klein reduction of the Kerr--Schild ansatz}
We now consider the KK reduction of the KS ansatz of the $\mathit{O}(D,D)$ generalised metric in $\mathcal{M}_{2D}$ to $\mathcal{N}_{2d}$.  The KS ansatz of the generalized metric is written as a sum of two parts, a background that solves the equations of motion and a fluctuation piece, which does not have to be small.  It is linear in the fluctuations yet exact without requiring any approximations.  Thus it is not an approximation but an exact solution of the full equations of motion. The fluctuation part can be interpreted as fields on the KS background, as in metric perturbation theory. The crucial property of the KS ansatz is that it reduces the nonlinear equations of motion to a set of linear partial differential equations.

Let us assume that the background and total KS space are both described by doubled torus bundles. We introduce an arbitrary background generalised metric $\hat{\mathcal{H}}_{0}$ on the doubled torus bundle patch, which does not have to be flat. The KS ansatz for the $\mathit{O}(D,D)$ generalised metric $\hat{\mathcal{H}}$ is given by \cite{Lee:2018gxc}
\begin{equation}
  \hat{\mathcal{H}}_{\hat{M}\hat{N}} = \hat{\mathcal{H}}_{0\hat{M}\hat{N}} + \kappa \varphi\big(\hat{K}_{\hat{M}} \hat{\bar{K}}_{\hat{N}} + \hat{K}_{\hat{N}} \hat{\bar{K}}_{\hat{M}}\big)\,,
\label{KS_total}\end{equation}
where $\varphi$ is an $\mathit{O}(D,D)$ scalar field, and $\kappa$ is a parameter denoting the order of perturbation. The fluctuation part on the RHS is linear but finite. Thus it is not a linearised approximation. Further, this ansatz is invariant under the rescaling
\begin{equation}
  \varphi \to e^{\alpha-\beta}\varphi\,, \qquad \hat{K} \to e^{-\alpha} \hat{K}\,, \qquad \hat{\bar{K}} \to e^{\beta} \hat{\bar{K}}\,,
\label{}\end{equation}
where $\alpha$ and $\beta$ are constants.

Here $\hat{K}$ and $\hat{\bar{K}}$ are $\mathit{O}(D,D)$ null vectors, $\hat{K}_{\hat{M}} \hat{\mathcal{J}}^{\hat{M}\hat{N}} \hat{K}_{\hat{N}} = \hat{\bar{K}}_{\hat{M}} \hat{\mathcal{J}}^{\hat{M}\hat{N}} \hat{\bar{K}}_{\hat{N}} = 0$, satisfying the background chirality condition,
\begin{equation}
  \hat{P}_{0\hat{M}}{}^{\hat{N}} \hat{K}_{\hat{N}} = \hat{K}_{\hat{M}} \,,
  \qquad
  \hat{\bar{P}}_{0\hat{M}}{}^{\hat{N}} \hat{\bar{K}}_{\hat{N}} = \hat{\bar{K}}_{\hat{M}} \,,
\label{chiralityhatKKbar}\end{equation}
where $\hat{P}_{0}$ and $\hat{\bar{P}}_{0}$ are a pair of background projectors defined by
\begin{equation}
  \hat{P}_{0\hat{M}\hat{N}} = \frac{1}{2} \big(\hat{\mathcal{J}}_{\hat{M}\hat{N}} + \hat{\mathcal{H}}_{0\hat{M}\hat{N}}\big)\,,
  \qquad
  \hat{\bar{P}}_{0\hat{M}\hat{N}} = \frac{1}{2} \big(\hat{\mathcal{J}}_{\hat{M}\hat{N}} - \hat{\mathcal{H}}_{0\hat{M}\hat{N}}\big)\,.
\label{projectors}\end{equation}
These satisfy the defining properties of projection operators: $\hat{P}_{0}^{2} = \hat{P}_{0}$, $\hat{\bar{P}}_{0}^{2} = \hat{\bar{P}}_{0}$ and $\hat{P}_{0} \hat{\bar{P}}_{0} = 0$. Thus it is clear that $\hat{K}$ and $\hat{\bar{K}}$ are orthogonal to each other due to the fact that they have opposite chiralities.  As the projection operators are the squares of the double-vielbeins \eqref{defining_VVbar}, we call $\hat{K}$ and $\hat{\bar{K}}$ \emph{left} and \emph{right} null vectors, respectively. The KS ansatz exactly satisfies the $\mathit{O}(D,D)$ constraint \eqref{ODDconstraint} without any assumption or approximation.

We may solve the background chirality condition \eqref{chiralityhatKKbar} and determine the parametrisation of $\hat{K}$ and $\hat{\bar{K}}$ in terms of undoubled $D$-dimensional null vectors, $\hat{l}$ and $\hat{\bar{l}}$:
\begin{equation}
  \hat{K}_{\hat{M}} = \frac{1}{\sqrt{2}} \begin{pmatrix} \hat{l}^{\hat{\mu}} \\ \big(\hat{g}_{0\hat{\mu}\hat{\nu}} + \hat{B}_{0\hat{\mu}\hat{\nu}}\big) \hat{l}^{\hat{\nu}}\end{pmatrix}\,,
  \qquad
  \hat{\bar{K}}_{\hat{M}} = \frac{1}{\sqrt{2}} \begin{pmatrix} \hat{\bar{l}}^{\hat{\mu}} \\ \big(-\hat{g}_{0\hat{\mu}\hat{\nu}} + \hat{B}_{0\hat{\mu}\hat{\nu}}\big) \hat{\bar{l}}^{\hat{\nu}}\end{pmatrix}\,,
\label{parahatKKbar}\end{equation}
where $\hat{g}_{0}{}_{\hat{\mu}\hat{\nu}}$ and $\hat{B}_{0}{}_{\hmu\hnu}$ are the background metric and background Kalb-Ramond field, respectively. The null conditions for $\hat{K}$ and $\hat{\bar{K}}$ imply $\hat{l}$ and $\hat{\bar{l}}$ are also null with respect to the background metric $\hat{g}_{0}$,
\begin{equation}
  \hat{l}^{\hat{\mu}} \hat{g}_{0\hat{\mu}\hat{\nu}} \hat{l}^{\hat{\nu}} = 0 \,, \qquad \hat{\bar{l}}^{\hat{\mu}} \hat{g}_{0\hat{\mu}\hat{\nu}} \hat{\bar{l}}^{\hat{\nu}} = 0 \,.
\label{}\end{equation}
However, they do not have to be mutually orthogonal, so in general $\hat{l}^{\hat{\mu}}\hat{g}_{0\hat{\mu}\hat{\nu}} \hat{\bar{l}}^{\hat{\nu}} \neq 0$. If we further impose an orthogonality condition between them, it follows that they must be proportional to each other. We also refer to $\hat{l}$ and $\hat{\bar{l}}$ as left and right null vectors, analogously to $\hat{K}$ and $\hat{\bar{K}}$.

Let us consider the KK reduction of the KS ansatz \eqref{KS_total}. Note that we can recast the parametrisation of $\hat{K}$ and $\hat{\bar{K}}$ in \eqref{parahatKKbar} by using the background double-vielbein,
\begin{equation}
  \hat{K}_{\hat{M}} = \hat{V}_{0\hat{M}}{}^{\hat{m}} \hat{l}_{\hat{m}}\,, \qquad   \hat{\bar{K}}_{\hat{M}} = \hat{\bar{V}}_{0\hat{M}}{}^{\hat{\bar{m}}} \hat{\bar{l}}_{\hat{\bar{m}}} \,,
\label{K_DV}\end{equation}
where $\hat{l}_{\hat{m}} = \hat{e}_{0\hat{m}}{}^{\hat{\mu}} \hat{l}_{\hat{\mu}}$ and $\hat{\bar{l}}_{\hat{m}} = \hat{\bar{e}}_{0\hat{\bar{m}}}{}^{\hat{\mu}} \hat{\bar{l}}_{\hat{\mu}}$, and $\hat{e}_{0}$ and $\hat{\bar{e}}_{0}$ are background vielbeins associated to the common background metric $\hat{g}_{0}$. This expression is useful in determining the KK ansatz of $\hat{K}$ and $\hat{\bar{K}}$. First, we assume that the $D$-dimensional vectors are decomposed as
\begin{equation}
  \hat{l}_{\hat{m}} = \big(l_{m} \,, j_{a}\big) \,,
  \qquad
  \hat{\bar{l}}_{\hat{\bar{m}}} = \big(\bar{l}_{m} \,, \bar{j}_{a}\big) \,.
\label{KK_llb}\end{equation}
Substituting the KK ansatz for the double-vielbeins \eqref{KK_DV} and the $D$-dimensional null vectors \eqref{KK_llb} into \eqref{K_DV}, we obtain the KK ansatz for $\hat{K}$ and $\hat{\bar{K}}$,
\begin{equation}
\begin{aligned}
  \hat{K}_{\hat{M}} =  \begin{pmatrix}
    K_{M} - \frac{1}{2} W_{0M}{}^{A} (W^{t}_{0})_{A}{}^{N} K_N - W_{0M}{}^{A} J_{A}
    \\
    J_{A} + (W_{0}^{t})_{A}{}^{M} K_{M}
  \end{pmatrix}\,,
  \\
  \hat{\bar{K}}_{\hat M} =\begin{pmatrix}
  \bar{K}_{M} - \frac{1}{2} W_{0M}{}^{A} (W^{t}_{0})_{A}{}^{N} \bar{K}_{N} - W_{0M}{}^{A} \bar{J}_{A}
  \\
  \bar{J}_{A} + (W_{0}^{t})_{A}{}^{M} \bar{K}_{M}
  \end{pmatrix}\,,
\end{aligned}
\label{smaller_doubled_vectors}\end{equation}
where $K_{M}$ and $\bar{K}_{M}$ are $\mathit{O}(d,d)$ vectors, and $K_{A}$ and $\bar{K}_{A}$ are $\mathit{O}(n,n)$ vectors. These also satisfy the chirality conditions with respect to the $\mathit{O}(d,d)$ and $\mathit{O}(n,n)$ projection operators, respectively. Similar to \eqref{parahatKKbar}, we can parametrise $K$ and $\bar{K}$ using \eqref{KK_llb} as
\begin{equation}
\begin{aligned}
   K_{M} &= V_{0M}{}^{m} l_{m} = \frac{1}{\sqrt{2}} \begin{pmatrix}
    l^\mu \\ (g_{0} +B_{0})_{\mu\nu} l^\nu
  \end{pmatrix}\,,
  &\qquad
  \bar{K}_{M} &= \bar{V}_{0M}{}^{\bar{m}} \bar{l}_{\bar{m}} = \frac{1}{\sqrt{2}}\begin{pmatrix}
    \bar{l}^\mu \\ (-g_{0}+B_{0})_{\mu\nu} \bar{l}^\nu
  \end{pmatrix} \,,
  \\
  J_{A} &= V_{0A}{}^{a} j_{a} = \frac{1}{\sqrt{2}} \begin{pmatrix}
    j^\alpha \\	
    (h_{0}+b_{0})_{\alpha\beta} j^\beta
  \end{pmatrix}\,,
  &\qquad
  \bar{J}_{A} &= \bar{V}_{0A}{}^{\bar{a}} \bar{j}_{\bar{a}} = \frac{1}{\sqrt{2}}  \begin{pmatrix}
    \bar{j}^\alpha \\
    (-h_{0}+b_{0})_{\alpha\beta} \bar{j}^\beta
  \end{pmatrix}\,.
\end{aligned}\label{}
\end{equation}
It is important to note that $K$ and $\bar{K}$ are non-null $\mathit{O}(d,d)$ vectors in general (similarly $J_{A}$ and $\bar{J}_{A}$); rather, the null conditions simply decompose as
\begin{equation}
\begin{aligned}
  \hat{K}_{\hat{M}} \hat{K}^{\hat{M}} = K_{M} K^{M} + J_{A}J^{A} = 0 \,,
  \\
  \hat{\bar{K}}_{\hat{M}} \hat{\bar{K}}^{\hat{M}} = \bar{K}_{M} \bar{K}^{M} + \bar{J}_{A} \bar{J}^{A} = 0 \,.
\end{aligned}\label{}
\end{equation}

Since the parametrisation of $W_{MA}$ \eqref{para_W} is of the same form as the derivative operator satisfying the section condition,
\begin{equation}
  \partial_{M} = \begin{pmatrix} 0 \\ \partial_{\mu} \end{pmatrix}\,,
\end{equation}
we cannot represent the KS ansatz for $W_{MA}$ in terms of $K$ and $\bar{K}$. Instead, we introduce a pair of $\mathit{O}(d,d)$ null vectors $L_{M}$ and $\bar{L}_{M}$ satisfying the section condition,
\begin{equation}
  L^M L_M =\bar{L}^M \bar{L}_M = \bar{L}^M L_M=0\,,
\label{}\end{equation}
and require further that these are orthogonal to the derivative operator $\partial_{M}$: $L^{M} \partial_{M} = \bar{L}^{M} \partial_{M} = 0$.
It follows that the parametrisation of $L$ and $\bar{L}$ are the same as that of $\partial_{M}$,
\begin{equation}
  L_{M} = \frac{1}{\sqrt{2}} \begin{pmatrix} 0 \\ l_\mu \end{pmatrix},
  \qquad
  \bar{L}_M = \frac{1}{\sqrt{2}} \begin{pmatrix} 0 \\  \bar{l}_\mu \end{pmatrix} \,.
\label{}\end{equation}

Thus we obtain the KS ansatz for the KK decomposition of the $\mathit{O}(D,D)$ generalised metric,
\begin{equation}
\begin{aligned}
  W_{M}{}^{A} &= W_{0M}{}^{A} + \frac{2\varphi'}{1-\varphi'^{2}(l\cdot l)(\bar{l} \cdot \bar{l})} \Big( L_M \bar{J}^{A} +  \bar{L}_M J^{A} - \varphi' (\bar{l} \cdot \bar{l}) L_M J^{A} -\varphi' (l\cdot l) \bar{L}_M \bar{J}^{A} \Big) \, ,
  \\
  \mathcal{M}_{AB} &=  \mathcal{M}_{0AB} +\frac{2\varphi'}{1-\varphi'^{2}(l\cdot l)(\bar{l} \cdot \bar{l})} \Big(J_{A} \bar{J}_{B} + \bar{J}_{A} J_{B}
  -\varphi' (\bar{l} \cdot \bar{l}) J_{A} J_{B} - \varphi' (l\cdot l) \bar{J}_{A} \bar{J}_{B} \Big) \, ,
  \\
  \mathcal{H}_{MN} &= U_{M}{}^{P} \big(\hat{\mathcal{H}}_{0PQ} + \varphi (\hat{K}_{P} \hat{\bar{K}}_{Q} +\hat{K}_{Q} \hat{\bar{K}}_{P}) - W_{P}{}^{A} \mathcal{M}_{A B} (W^{t})^{B}{}_{Q} \big) (U^{t})^{Q}{}_{N}\,,
\end{aligned}\label{KS_ansatz_KK_reduction}
\end{equation}
where
\begin{equation}
  U_{M}{}^{P} = \delta_{M}{}^{P} +\frac{1}{2} W_{M}{}^{A} (W^t)_{A}{}^{P} \,,
  \qquad
  \varphi' = \frac{\frac{1}{2}\kappa\varphi}{1+\frac{\kappa\varphi}{2} (l\cdot \bar{l})}\,.
\label{varphip}\end{equation}
Similarly, from \eqref{smaller_doubled_vectors} we find that
\begin{equation}
\begin{aligned}
  \hat{K}_{M} &= K_{M} - \frac{1}{2} W_{0M}{}^{A} (W^{t}_{0})_{A}{}^{N} K_N - W_{0M}{}^{A} J_{A} \, ,
  \\
  \hat{\bar{K}}_{M} &= \bar{K}_{M} - \frac{1}{2} W_{0M}{}^{A} (W^{t}_{0})_{A}{}^{N} \bar{K}_{N} - W_{0M}{}^{A} \bar{J}_{A} \, .
\end{aligned}\label{}
\end{equation}
As a consistency check, we may turn off $J$ and $\bar{J}$, such that $K$ and $\bar{K}$ also become null vectors. In this case, $\mathcal{M}_{AB}$ and $W_{MA}$ vanish and \eqref{KS_ansatz_KK_reduction} reduces to the standard KS ansatz for the $\mathit{O}(d,d)$ generalised metric, as expected. Note that the lower-dimensional KS ansatz \eqref{KS_ansatz_KK_reduction} is nonlinear in $\kappa$, which counts the order of fluctuations because $K$ and $\bar{K}$ are not null in general. Remarkably, as we will see later, the corresponding equations of motion are nevertheless linear even after KK reduction.

We now construct the lower-dimensional KS ansatzes for the supergravity fields using the parametrization of $\hat{\mathcal{H}}$ \eqref{para_hatH}. Before KK reduction, the KS ansatz is given by
\begin{equation}
\begin{aligned}
  \hat g^{\hat \mu \hat\nu} &= \hat{g}_{0}^{\hat\mu\hat\nu} + \frac{\kappa\varphi}{2} \left(\hat{l}^{\hat\mu} \hat{\bar{l}}^{\hat\nu} + \hat{l}^{\hat \nu} \hat{\bar{l}}^{\hat \mu}\right)\,,
  \\
  \hat{g}_{\hat \mu\hat \nu} &= \hat{g}_{0\hat\mu \hat\nu} - \frac{1}{2} \frac{\kappa \varphi}{1+\frac{1}{2} \kappa \varphi( \hat{l}\cdot \hat{\bar{l}})} \left(\hat{l}_{\hat \mu} \hat{\bar{l}}_{\hat \nu} + \hat{l}_{\hat\nu} \hat{\bar{l}}_{\hat \mu}\right)\,,
  \\
  \hat{B}_{\hat\mu\hat\nu} &= \hat{B}_{\hat{\mu}\hat{\nu}} + \frac{1}{2} \frac{\kappa\varphi}{1+\frac{1}{2} \kappa \varphi(\hat{l} \cdot \hat{\bar{l}})} \left(\hat{l}_{\hat\mu} \hat{\bar{l}}_{\hat \nu} - \hat{l}_{\hat\nu} \hat{\bar{l}}_{\hat\mu}\right) \, .
\end{aligned}\label{KS_hghB}
\end{equation}
The KK ansatz of the null vectors $\hat{l}$ and $\hat{\bar{l}}$ are as follows:
\begin{equation}
\begin{aligned}
  \hat{l}_{\hat{\mu}} &= \begin{pmatrix} l_{\mu} + A_{0\mu}{}^{\alpha} j_{\alpha} \\ j_{\alpha} \end{pmatrix}\,,
    \qquad
  \hat{l}^{\hat{\mu}}  = \begin{pmatrix}
    l^{\mu} \\ j^\alpha - h_{0}^{\alpha\gamma} A_{0\gamma}{}^{\nu} l_{\nu} \end{pmatrix}\,,
  \\
 \hat{\bar{l}}_{\hat{\mu}} &= \begin{pmatrix} \bar{l}_{\mu} + A_{0\mu}{}^{\alpha} \bar{j}_{\alpha} \\ \bar{j}_{\alpha} \end{pmatrix}\,,
 \qquad
 \hat{\bar{l}}^{\hat{\mu}} = \begin{pmatrix} \bar{l}^{\mu} \\ \bar{j}^{\alpha} - h_{0}^{\alpha\gamma} A_{0\gamma}{}^{\nu} \bar l_{\nu}
  \end{pmatrix}\,.
\end{aligned}\label{KK_null}
\end{equation}
Using the Kaluza--Klein ansatz for $\hat{g}$ and $\hat{B}$ in \eqref{KKansatz_gBphi}, one can read off the KS ansatz for the lower-dimensional fields:
\begin{equation}
\begin{aligned}
   h_{\alpha\beta} &= h_{0\alpha\beta} - \frac{\varphi'}{1 + \varphi'(j\cdot \bar{j})} (j_\alpha \bar{j}_\beta + \bar{j}_\alpha j_\beta) \,,
   \\
   b_{\alpha\beta} &= b_{0\alpha\beta} + \frac{\varphi'}{1 + \varphi'(j\cdot \bar{j})}(j_\alpha \bar{j}_\beta - j_\beta \bar{j}_\alpha) \,,
   \\
   g^{\mu\nu} &= g_{0}^{\mu\nu} + \frac{\kappa \varphi}{2} \big(l^{\mu} \bar{l}^{\nu} + \bar{l}^{\mu} l^{\nu}\big)\,,
   \\
   g_{\mu\nu} &= g_{0\mu\nu}  - \frac{\varphi'}{1-\varphi'^{2}(l\cdot l)(\bar{l} \cdot \bar{l})}\left( l_\mu \bar{l}_\nu + \bar{l}_\mu l_\nu - \varphi' (\bar{l} \cdot \bar{l}) l_\mu l_\nu - \varphi' (l\cdot l) \bar{l}_\mu \bar{l}_\nu \right) \,,
   \\
   B_{\mu\nu} &= B_{0\mu\nu} + \frac{2\varphi'}{1-\varphi'^{2}(l\cdot l)(\bar{l} \cdot \bar{l})} l_{[\mu} \bar{l}_{\nu]} - \mathfrak{a}_{[\mu}{}^{\alpha} b_{0|\alpha\beta|}  (A_{0}^{t})^{\beta}{}_{\nu]} - A_{0[\mu}{}^{\alpha} (\mathfrak{c}^{t})_{|\alpha|\nu]} +\mathfrak{a}_{[\mu}{}^{\gamma} (C_{0}^{t})_{|\gamma|\nu]}
    \\
    A_{\mu}{}^{\alpha} &= A_{0\mu}{}^{\alpha} +\mathfrak{a}_{\mu}{}^{\alpha}\,,
    \\
    C_{\mu\alpha} &= C_{0\mu\alpha} + \mathfrak{c}_{\mu\alpha} - \mathfrak{a}_{\mu}{}^{\gamma} b_{0\gamma\alpha} \,,
\end{aligned}\label{KK_KS_undoubled}
\end{equation}
where
\begin{equation}
\begin{aligned}
  \mathfrak{a}_{\mu}{}^{\alpha} &= -\frac{\varphi'}{1-\varphi'^{2}(l\cdot l)(\bar{l} \cdot \bar{l})} \left(l_{\mu} \bar{j}^{\alpha} + \bar{l}_{\mu} j^{\alpha} -\varphi' (\bar{l} \cdot \bar{l}) l_{\mu} j^{\alpha}- \varphi' (l\cdot l)\bar{l}_{\mu} \bar{j}^{\alpha} \right) \,,
  \\
  \mathfrak{c}_{\mu \alpha} &= \frac{\varphi'}{1-\varphi'^{2}(l\cdot l)(\bar{l} \cdot \bar{l})} \left( l_{\mu} \bar{j}_{\alpha} -\bar{l}_{\mu} j_{\alpha} +\varphi' (\bar{l} \cdot \bar{l}) l_{\mu} j_{\alpha}- \varphi' (l\cdot l)\bar{l}_{\mu} \bar{j}_{\alpha} \right) \,.
\end{aligned}\label{}
\end{equation}

Similarly to $K$ and $\bar{K}$, $l$ and $\bar{l}$ are non-null vectors unless $j$ and $\bar{j}$ are absent. Such a relaxation of the null condition is reminiscent of the KS ansatz for heterotic DFT \cite{Cho:2019ype}. Let us consider this from the perspective of partial dimensional reduction. The local structure group of heterotic DFT is given by $\mathit{O}(1,D-1)_{L} \times \mathit{O}(1,D-1+n)_{R}$, and the KS ansatz consists of a $D$-dimensional null vector in the left (chiral) sector and a KK reduction of a $(D+n)$-dimensional null vector to a $D$-dimensional space in the right (antichiral) sector. From the $D$-dimensional point of view, the left vector is null but the right vector is not null. Thus the null condition is partially relaxed; however, the KS field equations for heterotic DFT are still linear. Note that if we set $j=0$ but $\bar{j}\neq 0$, \eqref{KK_KS_undoubled} indeed becomes the KS ansatz for heterotic DFT.

\subsection{KK reduction of the KS equations of motion}\label{sec2.3}
Our next goal is to show that the KS equations of motion of the ungauged half-maximal supergravities are linear. We assume the background is a flat torus bundle with a flat background metric $\hat{g}_{0}$ and ignore the background Kalb--Ramond field and dilaton. For a consistent KK reduction, we need to specify the torus coordinate explicitly in order to turn off the torus-coordinate dependence. To this end, we represent the KS ansatz in a bundle patch $\hat{U}$, which is given as a direct product between a base and a torus patch for the background torus bundle.

On the other hand, we may have a coordinate patch in which the background metric is trivial, $\hat{g}_{0} = \hat{\eta}$. We call it the ``Kerr--Schild patch'' and denote it as $\hat{U}_{\text{KS}}$. However, the KS patch cannot be a bundle patch $\hat{U}$ for nontrivial bundles due to their background bundle connection $A_{0}$, which is encoded in $\hat{g}_{0}$. In general, $\hat{U}_{KS}$ and $\hat{U}$ are related by a diffeomorphism, which maps the trivial metric in $\hat{U}_{KS}$ to a nontrivial flat metric $\hat{g}_{0}$ admitting $A_{0}$. In this case, the background connection is nonvanishing, and we need to exploit the background covariant derivatives $\hat{\nabla}_{0}$ to describe the KS equations.

As in GR, the equations of motion for DFT are determined by curvatures \cite{Jeon:2010rw,Jeon:2011cn}: the generalised Ricci tensor $\hat{\mathcal{R}}_{\hat{\mu}\hat{\nu}}$ and the generalised Ricci scalar $\hat{\mathcal{R}}$. These provide the equations of motion for the generalised metric and the DFT dilaton, respectively, and are equivalent to the equations of motion of NSNS supergravity. Note that $\hat{\mathcal{R}}_{\hat{\mu}\hat{\nu}}$ is neither symmetric nor antisymmetric: the symmetric part gives the equations of motion for the metric, while the antisymmetric part provides the equations for the Kalb--Ramond field. There may be additional constraints on the null vectors arising from the equations of motion, the so-called on-shell constraints \cite{Lee:2018gxc,Cho:2019ype}. These arise from the contraction between null vectors and the generalised Ricci tensor, $\hat{l}^{\hat{\mu}} \hat{\bar{l}}^{\hat{\nu}} \hat{\mathcal{R}}_{\hat{\mu}\hat{\nu}}=0$, giving
\begin{equation}
  \hat{l}^{\hat{\mu}} \hat{\nabla}_{0\hat{\mu}} \hat{\bar{l}}^{\nu} = 0\,, \qquad \hat{\bar{l}}^{\hat{\mu}} \hat{\nabla}_{0\hat{\mu}} \hat{l}^{\nu} = 0\,.
\label{on-shell}\end{equation}
Substituting the KS ansatz into the $\hat{\mathcal{R}}_{\hat{\mu}\hat{\nu}}$ and $\hat{\mathcal{R}}$ and applying the on-shell constraints, we obtain a set of linear equations:
\begin{equation}
\begin{aligned}
  \hat{\mathcal{R}} &= \hat{\nabla}_{0\hat{\mu}} \hat{\nabla}_{0\hat{\nu}}\left(\varphi \hat{l}^{\mu} \hat{\bar{l}}^{\nu}\right) -4 \hat{\nabla}_{0\hat{\mu}} \hat{\nabla}_{0}^{\hat{\mu}} f^{{\scriptscriptstyle(0)}}=0 \, ,
  \\
  \hat{\mathcal{R}}_{\hat{\mu}\hat{\nu}} &= \hat{\nabla}_{0}^{\hat{\rho}} \hat{\nabla}_{0\hat{\rho}}\left(\varphi \hat{l}_{\hat{\mu}} \hat{\bar{l}}_{\hat{\nu}}\right)
  - \hat{\nabla}_{0}^{\hat{\rho}} \hat{\nabla}_{0\hat{\mu}}\left(\varphi \hat{l}_{\hat{\rho}} \hat{\bar{l}}_{\hat{\nu}}\right)
  -\hat{\nabla}_{0}^{\hat{\rho}} \hat{\nabla}_{0\hat{\nu}} \left(\varphi \hat{l}_{\hat{\mu}} \hat{\bar{l}}_{\hat{\rho}}\right)
  +4\hat{\nabla}_{0\hat{\mu}} \hat{\nabla}_{0\hat{\nu}} f^{{\scriptscriptstyle(0)}}=0\,,
\end{aligned}\label{}
\end{equation}
where $f^{{\scriptscriptstyle(0)}}$ is the lowest term in the $\kappa$-expansion of $f$, $f = \sum_{n=0} f^{{\scriptscriptstyle(n)}} \kappa^{n}$.

The linearity of the higher-dimensional equations persists even after KK reduction because the Kaluza--Klein ansatzes for the null vectors do not carry any $\kappa$ dependence. We introduce two classes of compactification in accordance with the topology of the background geometry: trivial reduction and twisted reduction.

\subsubsection{Twisted reduction}
For a nontrivial flat background torus bundle, the KS coordinates cannot be identified with the bundle coordinates due to the presence of the bundle connection. To perform a KK reduction, we must map the KS patch $\hat{U}_{KS}$ to a bundle patch $\hat{U}$ via a diffeomorphism, such that the torus fibre coordinates appear explicitly. Then the trivial background metric $\hat{\eta}$ in $\hat{U}_{KS}$ is mapped to $\hat{g}_{0}$, which is a flat metric in $\hat{U}$. According to the KK ansatz, the transformed metric $\hat{g}_{0}$ is decomposed into
\begin{equation}
  \hat{\eta} \underset{\text{diffeo.}} \longrightarrow \hat{g}_{0} = \begin{pmatrix} g_{0\mu\nu} + A_{0} h_{0} (A_{0}^{t}) & A_{0}h_{0} \\ h_{0} (A_{0}^{t}) & h_{0} \end{pmatrix} \,.
\label{}\end{equation}
On the bundle patch $\hat{U}$, we can fix the coordinates so that the metric $\hat{g}_{0}$ is independent of the torus fibre coordinates $y^{\alpha}$ of the KK reduction,
\begin{equation}
  \hat{g}_{0} (x,y)= \hat{g}_{0}(x)\,.
\label{fibre_isometry}\end{equation}
Since $\hat{g}_{0}$ is flat, \eqref{fibre_isometry} implies that $h_{0}$ is a constant matrix, and $A_{0}^{\alpha}$ is a flat connection satisfying $F_{0}^{\alpha} = dA_{0}^{\alpha} = 0$. If we compute the spin connection using the KK ansatz for $\hat{g}_{0}$, the only nonvanishing components are the external part,
\begin{equation}
  \begin{cases}
  	\hat{\omega}_{0\bar{p}mn} = (e_{0}^{-1})_{m}{}^{\mu}\omega_{0\mu mn} \, ,
  	\\
  	0 \qquad \text{otherwise} \, .
  \end{cases}
\label{}\end{equation}

Thus the on-shell constraints reduce to
\begin{equation}
  l^{\mu} \nabla_{0\mu} \bar{l}^{\nu} = 0\,, \qquad \bar{l}^{\mu} \nabla_{0\mu} l^{\nu} = 0\,, \qquad l^{\mu} \partial_{\mu} \bar{j}^{\alpha} = 0\,, \qquad \bar{l}^{\mu} \partial_{\mu} j^{\alpha} =0\,.
\label{}\end{equation}
Similarly, the KS equations of motion reduce to
\begin{equation}
\begin{aligned}
  \hat{\mathcal{R}}_{\mu\nu} &= \nabla_{0}^{\rho}\nabla_{0\rho}\left(\varphi l_{\mu} \bar{l}_{\nu}\right)-\nabla_{0}^{\rho} \nabla_{0\mu}\left(\varphi l_{\rho} \bar{l}_{\nu}\right)-\nabla_{0}^{\rho} \nabla_{0\nu} \left(\varphi l_{\mu} \bar{l}_{\rho}\right) + 4\nabla_{0 \mu} \partial_{\nu} f^{{\scriptscriptstyle(0)}} = 0 \,,
  \\
  \hat{\mathcal{R}}_{\mu\alpha} &= \nabla_{0}^{\rho}\nabla_{0\rho}\left(\varphi l_{\mu} \bar{j}_{\alpha}\right)-\nabla_{0}^{\rho} \nabla_{0\mu}\left(\varphi l_{\rho} \bar{j}_{\alpha}\right) = 0\,,
  \\
  \hat{\mathcal{R}}_{\alpha\mu} &= \nabla_{0}^{\rho}\nabla_{0\rho}\left(\varphi j_{\alpha} \bar{l}_{\mu}\right)-\nabla_{0}^{\rho} \nabla_{0\mu}\left(\varphi j_{\alpha} \bar{l}_{\rho}\right) = 0 \,,
  \\
  \hat{\mathcal{R}}_{\alpha\beta} &= \nabla_{0}^{\rho} \partial_{\rho} \big(\varphi j_{\alpha} \bar{j}_{\beta}\big) = 0 \,.
\end{aligned}\label{KS_EOM_KK_reduction}
\end{equation}
It is remarkable that these equations are completely linear even though the KK reduction of the KS ansatz in \eqref{KK_KS_undoubled} is highly nonlinear. This is because $\hat{\mathcal{R}}_{\mu\nu}$ provides not only the equations of motion of the external metric $g$ but also a combination of the $\mathit{U}(1)$ gauge field and scalar equations of motion. For example, the variation of metric $\hat{g}_{\mu\nu}$ is given, according to the KK ansatz, by
\begin{equation}
  \delta \hat{g} = \delta g + \delta \big(A h A^{t}\big)\,.
\end{equation}
This implies that while the equations of motion of $g_{\mu\nu}$ itself are nonlinear, some specific linear combinations of the equations of motion are in fact linear.

\subsubsection{Trivial reduction}
For a trivial bundle we can simply identify the KS patch and the bundle patch. Then the metric reduces to the trivial metric, $\hat{\eta}$, and is decomposed simply as
\begin{equation}
  \hat{\eta}_{\hat{\mu}\hat{\nu}} = \begin{pmatrix} \eta_{\mu\nu} & 0 \\ 0 & \delta_{\alpha\beta} \end{pmatrix} \,.
\label{}\end{equation}
Since the background connection $A_{0}$ vanishes, the KK ansatz of the null vectors are decomposed as simply
\begin{equation}
  \hat{l}_{\hat{\mu}} = \begin{pmatrix} l_{\mu} \\ j_{\alpha} \end{pmatrix}\,,
  \qquad
  \hat{\bar{l}}_{\hat{\mu}} = \begin{pmatrix} \bar{l}_{\mu} \\ \bar{j}_{\alpha} \end{pmatrix}\,.
\end{equation}
The KS equations for trivial bundles are obtained by replacing the background covariant derivative in \eqref{KS_EOM_KK_reduction} with partial derivatives, $\nabla_{0\mu} \to \partial_{\mu}$, giving
\begin{equation}
\begin{aligned}
  \hat{\mathcal{R}}_{\mu\nu} &= \partial^{\rho}\partial_{\rho}\left(\varphi l_{\mu} \bar{l}_{\nu}\right)-\partial^{\rho} \partial_{\mu}\left(\varphi l_{\rho} \bar{l}_{\nu}\right)-\partial^{\rho} \partial_{\nu} \left(\varphi l_{\mu} \bar{l}_{\rho}\right) + 4\partial_{\mu} \partial_{\nu} f^{{\scriptscriptstyle(0)}} = 0 \,,
  \\
  \hat{\mathcal{R}}_{\mu\alpha} &= \partial^{\rho}\partial_{\rho}\left(\varphi l_{\mu} \bar{j}_{\alpha}\right)-\partial^{\rho} \partial_{\mu}\left(\varphi l_{\rho} \bar{j}_{\alpha}\right) = 0\,,
  \\
  \hat{\mathcal{R}}_{\alpha\mu} &= \partial^{\rho}\partial_{\rho}\left(\varphi j_{\alpha} \bar{l}_{\mu}\right)-\partial^{\rho} \partial_{\mu}\left(\varphi j_{\alpha} \bar{l}_{\rho}\right) = 0 \,,
  \\
  \hat{\mathcal{R}}_{\alpha\beta} &= \partial^{\rho} \partial_{\rho} \big(\varphi j_{\alpha} \bar{j}_{\beta}\big) = 0 \,.
\end{aligned}
\end{equation}
%

\section{$\mathit{O}(D,D)$ transformation}\label{sec:3}
T-duality is the most distinguishing characteristic feature of half-maximal supergravities in a $T^{n}$-bundle with isometries. It is a powerful technique for generating solutions from a known seed solution, and many physically important solutions have been constructed this way \cite{Giveon:1994fu}. DFT is manifestly $\mathit{O}(D,D)$ covariant by construction and thus provides an efficient tool for describing T-duality. We will study how the KS ansatz transforms under a T-duality rotation, both for the basic Buscher rule and for a continuous $\mathit{O}(D,D)$ rotation.

\subsection{Buscher Rule}
The simplest T-duality transformation is the Buscher rule. It is derived from a string sigma model with an isometry in a curved target spacetime \cite{Buscher:1987qj,Rocek:1991ps}. If we denote the isometry direction as $z$, under a KS ansatz the Buscher rule for the background fields is
\begin{equation}
\begin{aligned}
&\tg'_{zz} = \frac{1}{\tg_{zz}} \,,\qquad \qquad  \tg'_{iz} = \frac{\tB_{iz}}{\tg_{zz}} \,, \qquad \qquad  \tB'_{iz} = \frac{\tg_{iz}}{\tg_{zz}}\,,
\\
&\tg'_{ij} = \tg_{ij} - \frac{\tg_{iz}\tg_{jz} - \tB_{iz}\tB_{jz}}{\tg_{zz}}\,, \qquad \tB'_{ij} = \tB_{ij} - \frac{\tB_{iz}\tg_{jz} - \tg_{iz}\tB_{jz}}{\tg_{zz}}\,,
\end{aligned}\label{Buscher_gB}
\end{equation}
and for the null vectors is \cite{Lee:2018gxc}
\begin{equation}
\begin{aligned}
  l'_{i} &= l_{i} - \frac{(\tg_{iz}-\tB_{iz})l_{z}}{\tg_{zz}}\,,\qquad l'_{z} = \frac{l_{z}}{\tg_{zz}}\,,
  \\
  \bar{l}'_{i} &= \bar{l}_{i} - \frac{(\tg_{iz}+\tB_{iz})\bar{l}_{z}}{\tg_{zz}}\,,\qquad \bar{l}'_{z} = -\frac{\bar{l}_{z}}{\tg_{zz}}\,.
\end{aligned}\label{Buscher_null}
\end{equation}

When we consider a trivial background, $g_{0} = \eta$ and $B_{0} = 0$, the Buscher rule reduces to sign-flipping transformations,
\begin{equation}
\begin{aligned}
  \mathrm{I} &: \big(l_{i}\,,j_{z}\,,\bar{l}_{i}\,,\bar{j}_{z}\,,\varphi\,,f^{(0)}\big) \to \big(l_{i}\,,-j_{z}\,,\bar{l}_{i}\,,\bar{j}_{z}\,,\varphi\,,- f^{(0)}\big) \, ,
  \\
  \mathrm{II} &: \big(l_{i}\,,j_{z}\,,\bar{l}_{i}\,,\bar{j}_{z}\,,\varphi\,,f^{(0)}\big) \to \big(l_{i}\,,j_{z}\,,\bar{l}_{i}\,,-\bar{j}_{z}\,,\alpha\varphi\,,- f^{(0)}\big) \, ,
  \\
  \mathrm{III} &: \big(l_{i}\,,j_{z}\,,\bar{l}_{i}\,,\bar{j}_{z}\,,\varphi\,,f^{(0)}\big) \to \big(l_{i}\,,-j_{z}\,,\bar{l}_{i}\,,-\bar{j}_{z}\,,\varphi\,, f^{(0)}\big) \, .
\end{aligned}\label{buscher_rule_KS}
\end{equation}
Compared with the full Buscher transformation, this prescription is remarkably simple and handy for generating new solutions.


\subsection{General $\mathit{O}(n,n)$ rotation} \label{GlobalOnn}
We now consider more general T-duality transformations along the $y^{\alpha}$ directions. In this case T-duality is realised by $\mathit{O}(n,n)$ rotations which are embedded into the $\mathit{O}(D,D)$ group as
\begin{equation}
\hcO_{\hat{M}}{}^{\hat{N}} = \begin{pmatrix} \delta_{M}{}^{N} & 0 \\ 0 & \cO_{A}{}^{B}\end{pmatrix} \, ,
\label{}\end{equation}
where $\cO_{A}{}^{B}\in \mathit{O}(n,n)$ and is denoted by
\begin{equation}
\cO_{A}{}^{B} = \begin{pmatrix} a^{\alpha}{}_{\beta} & b^{\alpha\beta} \\ c_{\alpha\beta} & d_{\alpha}{}^{\beta} \end{pmatrix}\,.
\label{Onn}\end{equation}
Here $a$, $b$, $c$ and $d$ are $n\times n$ matrices satisfying
\begin{equation}
a b^{t}+ba^{t} = 0\,,\qquad a d^{t} + b c^{t} = \mathbf{1}_{n}\,, \qquad cd^{t} + dc^{t} = 0\,.
\label{ODDconstraints}\end{equation}
Under $\mathit{O}(D,D)$ transformations the generalised metric $\hat{\mathcal{H}}$ and the doubled null vectors $\hat{K}$ and $\hat{\bar{K}}$ transform linearly as
\begin{equation}
  \hat{\mathcal{H}}_{\hat{M}\hat{N}} \to \hcO_{\hat{M}}{}^{\hat{P}} \hat{\mathcal{H}}_{\hat{P}\hat{Q}} \big(\hcO^{t}\big){}^{\hat{Q}}{}_{\hat{N}}\,,
  \qquad
  \hat{K}_{\hat{M}} \to \hcO_{\hat{M}}{}^{\hat{N}} \hat{K}_{\hat{N}}\,,
  \qquad
  \hat{\bar{K}}_{\hat{M}} \to \hcO_{\hat{M}}{}^{\hat{N}} \hat{\bar{K}}_{\hat{N}} \,.
\label{}\end{equation}
The $\mathit{O}(n,n)$ subgroup acts on the $\mathit{O}(n,n)$ vector indices, $A,B,\cdots$, and the component fields transform as
\begin{equation}
  \mathcal{M}'_{AB} = \cO_{A}{}^{C} \mathcal{M}_{CD} (\cO^{t})^{D}{}_{B}\,,
  \qquad
  W'_{MA} = W_{MB} (\cO^{t})^{B}{}_{A} \, ,
\label{Onn_gen_met}\end{equation}
and
\begin{equation}
\begin{aligned}
  K'_{M} &= K_{M}\,, \qquad &J'_{A} &= \cO_{A}{}^{B} J_{B}\,,
  \\
  \bar{K}'_{M} &= \bar{K}_{M}\,,\qquad &\bar{J}'_{A} &= \cO_{A}{}^{B} \bar{J}_{B}\,.
\end{aligned}\label{Onn_transf}
\end{equation}

The parametrisation of $J$ and $\bar{J}$ in terms of $d$-dimensional vectors is given by
\begin{equation}
\begin{aligned}
J_{A} = \frac{1}{\sqrt{2}} \begin{pmatrix} j^{\alpha} \\ (\th + \tb)_{\alpha\beta} j^{\beta} \end{pmatrix}\,, \qquad
\bar{J}_{A} = \frac{1}{\sqrt{2}} \begin{pmatrix} \bar{j}^{\alpha} \\ (-\th + \tb)_{\alpha\beta} \bar{j}^{\beta} \end{pmatrix}\,.
\end{aligned}\label{}
\end{equation}
Substituting \eqref{Onn} into \eqref{Onn_transf}, we obtain the associated transformations of $j$ and $\bar{j}$,
\begin{equation}
\begin{aligned}
J'{}_{A} &= \frac{1}{\sqrt{2}} \begin{pmatrix} j'^{\alpha} \\ (h'_{0} + b'_{0})_{\alpha\beta} j'^{\beta} \end{pmatrix} %
= \frac{1}{\sqrt{2}} \begin{pmatrix} a^{\alpha}{}_{\beta} j^{\beta} + b^{\alpha\gamma} (\th+\tb)_{\gamma\beta}j^{\beta} \\ c_{\alpha\beta} j^{\beta} + d_{\alpha}{}^{\gamma} (\th+\tb)_{\gamma\beta}j^{\beta}\end{pmatrix}\,,
\\
\bar{J}'{}_{A} &= \frac{1}{\sqrt{2}} \begin{pmatrix} \bar{j}'^{\alpha} \\ (-h'_{0} + b'_{0})_{\alpha\beta} \bar{j}'^{\beta} \end{pmatrix} %
= \frac{1}{\sqrt{2}} \begin{pmatrix} a^{\alpha}{}_{\beta} \bar{j}^{\beta} + b^{\alpha\gamma} (-\th+\tb)_{\gamma\beta}\bar{j}^{\beta} \\ c_{\alpha\beta} \bar{j}^{\beta} + d_{\alpha}{}^{\gamma} (-\th+\tb)_{\gamma\beta}\bar{j}^{\beta}\end{pmatrix}\,.
\end{aligned}\label{transf_J}
\end{equation}
We can easily read off $j'^{\alpha}$ and $\bar{j}'^{\alpha}$ from the upper block of the equations
\begin{equation}
\begin{aligned}
j'^{\alpha} = a^{\alpha}{}_{\beta} j^{\beta} + b^{\alpha\gamma} \tE_{\gamma\beta}j^{\beta}\,,\qquad \bar{j}'^{\alpha} = a^{\alpha}{}_{\beta} \bar{j}^{\beta} - b^{\alpha\gamma} \tE^{t}_{\gamma\beta} \bar{j}^{\beta}\,,\qquad
\end{aligned}\label{nullprime}
\end{equation}
where $\th_{\alpha\beta}+\tb_{\alpha\beta} = \tE_{\alpha\beta}$. Further, we can determine $E_{0}'$ from each lower block equation of \eqref{transf_J},
\begin{equation}
\begin{aligned}
E'_{0}{}_{\alpha\beta} &= (c + d\tE){}_{\alpha\gamma}\big((a + b \tE)^{-1}\big){}^{\gamma}{}_{\beta} \, ,
\\
E'_{0}{}^{t}{}_{\alpha\beta} &= (-c + d\tE^{t}){}_{\alpha\gamma}\big((a -b \tE^{t})^{-1}\big){}^{\gamma}{}_{\beta} \, .
\end{aligned}\label{EET}
\end{equation}
Interestingly, this result is consistent with the $\mathit{O}(n,n)$ transformation of the generalised metric in \eqref{Onn_gen_met}.

The transformation \eqref{EET} encompasses several different types of transformations \cite{Giveon:1994fu}.  
\begin{itemize}
	\item $GL(n)$ global diffeomorphisms: $b = 0$ and $c = 0$, which from \eqref{ODDconstraints} implies $a = (d^{t})^{-1}$.  The background transformation is simply
	\begin{equation}
	h'_{0} = d\th d^{t} \, , \qquad b'_{0} = d\tb d^{t} \, .
	\end{equation}
	\item B-field shifts: $a = \mathbf{1}_{n}$, $d = \mathbf{1}_{n}$ and $b = 0$, thus $c = -c^{t}$.  The transformation is simply
	\begin{equation}
	h'_{0} = \th \, , \qquad b'_{0} = \tb + c \, .
	\end{equation}
	\item T-duality in $k$ dimensions: $a = \mathrm{diag}\{\mathbf{1}_{n-k},0\} = (d^{t})^{-1}$, while $b = \mathrm{diag}\{0,\mathbf{1}_{k}\} = (c^{t})^{-1}$.  For $k = 1$ we recover the Buscher rules \eqref{Buscher_gB}.
\end{itemize}

In most cases, we wish to preserve the asymptotic structure after the duality rotation, such as asymptotic flatness conditions. To this end we may choose the maximally compact subgroup $\mathit{O}(n)\times \mathit{O}(n)$ or $\mathit{O}(1,n-1)\times \mathit{O}(1,n-1)$ which preserves the flat metric.  In order to understand how this subgroup is embedded into the general transformation, consider the case of a flat Lorentz metric $\eta$ and vanishing B-field.  
The null vectors transform as
\begin{equation}
\begin{aligned}
j'{}^{\alpha} &= \left(a^{\alpha}{}_{\beta} + b^{\alpha\gamma}\eta_{\gamma\beta}\right)j^{\beta} \equiv o_{1}{}^{\alpha}{}_{\beta}j^{\beta} \, , 
\\
\bar{j}'{}^{\alpha} &= \left(a^{\alpha}{}_{\beta} - b^{\alpha\gamma}\eta_{\gamma\beta}\right)\bar{j}^{\beta} \equiv o_{2}{}^{\alpha}{}_{\beta}\bar{j}^{\beta} \, ,
\end{aligned} \label{jjbarflatprime}
\end{equation}
where in the last step we have defined the $\mathit{O}(1,n-1)$ rotations
\begin{equation}
o_{1} \equiv a + b\eta \, , \qquad o_{2} \equiv a - b\eta \, . \label{o12}
\end{equation}
We may similarly define
\begin{equation}
o_{3} \equiv d + c\eta^{-1} \, , \qquad o_{4} \equiv d - c\eta^{-1} \, , \label{o34}
\end{equation}
such that \eqref{EET} may be written as
\begin{equation} \label{hbp}
h'_{0} + b'_{0} = o_{3}\eta(o_{1})^{-1} \,, \qquad h'_{0} - b'_{0} = o_{4}\eta(o_{2})^{-1} \, .
\end{equation}
Note that \eqref{nullprime} 
is the generalisation of 
\eqref{jjbarflatprime} 
to curved backgrounds; however, only the $o_i$ transformations are global $\mathit{O}(n,n)$ elements.

The $\mathit{O}(n,n)$ conditions \eqref{ODDconstraints} give  constraints on $a,b,c$ and $d$, which in turn imply non-trivial relations between $o_{1}$, $o_{2}$, $o_{3}$ and $o_{4}$.  Writing \eqref{ODDconstraints} in terms of \eqref{o12} and \eqref{o34}, we find
\begin{align}
o_{1}\eta^{-1}o_{1}^t &= o_{2}\eta^{-1}o_{2}^t \, , \label{onnab} \\
o_{3}\eta o_{3}^t &= o_{4}\eta o_{4}^t \, , \label {onncd} \\
\mathbf{1}_{n} &= \frac{1}{2}\left[o_{1}o_{3}^t + o_{2}o_{4}^t\right] \, . \label{onnadbc}
\end{align}
In addition, by comparing the first of \eqref{hbp} to its transpose we see that
\begin{equation}
(o_{1}^{t})^{-1}\eta o_{3}^t = o_{4}\eta(o_{2})^{-1} \, .
\end{equation}
This property can be verified explicitly using \eqref{ODDconstraints}.


The maximal compact subgroup corresponds to $n^{2} - n$ degrees of freedom.  However the full $\mathit{O}(n,n)$ transformation has $2n^2 - n$ independent parameters: the additional redundancy arises from the fact that we can redefine $b$ and $c$ up to constant shifts of the B-field.  To fully specify the global transformation we may, 
for example demand that $b_{0} = 0$ implies $b'_{0} = 0$.
Comparing with \eqref{hbp}, we see that this choice corresponds to identifying $o_{3} = (o_{1}^t)^{-1}$ and $o_{4} = (o_{2}^t)^{-1}$.  Of the $\mathit{O}(n,n)$ constraints, \eqref{onnadbc} is trivially satisfied while \eqref{onnab} and \eqref{onncd} are satisfied for independent $o_{1}$ and $o_{2}$ as a result of the Lorentz conditions,
\begin{equation}
\eta = o_{1}^t\eta o_{1} = o_{2}^t\eta o_{2} \, .
\end{equation}
Thus the embedding of the double Lorentz group in $\mathit{O}(n,n)$ is parametrized by
\begin{equation}
\cO = \frac{1}{2}\begin{pmatrix} o_{1} + o_{2} & \left(o_{1} - o_{2}\right)\eta^{-1} \\ \eta\left(o_{1} - o_{2}\right) & \eta\left(o_{1} + o_{2}\right)\eta^{-1} \end{pmatrix} \, . \label{Onn2xLorentz}
\end{equation}

Having obtained an expression for our $\mathit{O}(n,n)$ transformation, we may also apply it to curved backgrounds and cases with non-trivial background B-field.  Nevertheless, we will mostly be interested in situations where the initial background B-field vanishes, $\tb = 0$.  By expanding an $\mathit{O}(n,n)$ rotation of the background metric, we find in this case the transformation rules\footnote{For general expressions of $\mathit{O}(D,D)$ transformations on curved backgrounds, see Appendix \ref{GeneralOnn}.}
\begin{align}
h_{0}^{-1}{}' &= ah_{0}^{-1}a^t + bh_{0} b^t \, , \label{Onnginv} \\
b'_{0}h_{0}^{-1}{}' &= ch_{0}^{-1}a^t + dh_{0} b^t \, , \label{OnnBginv} \\
h'_{0} - b'_{0}h_{0}^{-1}{}'b'_{0} &= ch_{0}^{-1}c^t + dh_{0} d^t \, .
\end{align}




\section{Classical double copy and KK reduction} \label{sec:4}
The double copy is a relation between scattering amplitudes in gravity and Yang--Mills theory. It is closely related to the KLT relation \cite{Kawai:1985xq}, which arises from the left-right mover decomposition in the closed-string mode expansion. Each left and right mover corresponds to open-string modes. From the low-energy effective field theory point of view, the double copy relates (super)gravity to a pair of Yang--Mills theories. Recently it has been shown that DFT and the double copy share the same origin from string theory \cite{Lee:2018gxc,Cho:2019ype}. The double Lorentz groups \eqref{double_local_Lorentz} also reflect the left-right mode decomposition because each mode perceives spacetime independently. Furthermore, the double copy has recently been extended to classical solutions of the equations of motion \cite{Monteiro:2011pc} using the KS formalism. This is called the classical double copy, under which exact gravitational solutions of the KS type can be represented by solutions of the Maxwell theory. In this section we will discuss how to define the classical double copy for half-maximal supergravities and their T-duality rotation by exploiting the results of the previous sections.

The mode expansion of the closed string in a toroidal background also exhibits the left-right mode decomposition, as in a flat target spacetime \cite{Polchinski:1998rq}. We neglect the winding modes since they are suppressed in the low-energy effective field theory. Then the left and right movers correspond to mode expansions of open strings with the same KK reduction. This implies that the single copies for half-maximal supergravities are given by the KK reduction of a pair of higher-dimensional Yang--Mills theories,
\begin{equation}
\begin{aligned}
  \mathcal{L}_{\text{Left}} = \Tr\left[ -\frac{1}{4} F_{\mu\nu} F^{\mu\nu} - \frac{1}{2} D_{\mu} X^{\alpha} D^{\mu} X_{\alpha} + \big[X^{\alpha},X^{\beta}\big] \big[ X_{\alpha},X_{\beta}\big]\right]\,,
  \\
  \mathcal{L}_{\text{Right}} = \Tr\left[ -\frac{1}{4} \bar{F}_{\mu\nu} \bar{F}^{\mu\nu} - \frac{1}{2} \bar{D}_{\mu} \bar{X}^{\alpha} \bar{D}^{\mu} \bar{X}_{\alpha} + \big[\bar{X}^{\alpha},\bar{X}^{\beta}\big] \big[ \bar{X}_{\alpha},\bar{X}_{\beta}\big]\right]\,,
\end{aligned}\label{single_copy_action}
\end{equation}
where $F_{\mu\nu}$ and $\bar{F}_{\mu\nu}$ are the Yang--Mills field strengths for $A$ and $\bar{A}$ respectively, and $D_{\mu}$ and $\bar{D}_{\mu}$ are covariant derivatives defined by
\begin{equation}
  D_{\mu}\ \bullet = \partial_{\mu} \bullet + [A_{\mu}, \bullet ] \,,
  \qquad
  \bar{D}_{\mu} \ \bullet = \partial_{\mu} \bullet + [\bar{A}_{\mu}, \bullet ] \,.
\label{}\end{equation}

As we have discussed in Section \ref{sec2.3}, we must consider a bundle patch $\hat{U}$ that is given by a direct product of a patch of the base $U$ and of the torus fibre $T^{n}$. After KK reduction, the background metric $g_{0}$ remains flat but is still non-trivial in $U$. We assume that the base spacetime $N_{d}$ (including the background) admits a Killing vector $\xi$, and furthermore require that this Killing vector is covariantly constant in order to be able to put the Killing vector inside the background covariant derivative which appears in the KS equations of motion \eqref{KS_EOM_KK_reduction}. Using the properties of Killing vectors, one can show that the covariantly constant condition implies that $\xi_{\mu}$ is a closed form, $\mathrm{d}\xi =0$:
\begin{equation}
  \nabla_{0\mu} \xi^{\rho} = g_{0}^{\rho\nu} \nabla_{0\mu} \xi_{\nu} = \frac{1}{2} g_{0}^{\rho\nu}\big(\nabla_{0\mu} \xi_{\nu} - \nabla_{0\nu} \xi_{\mu}\big) = \frac{1}{2} g_{0}^{\rho\nu}\big(\partial_{\mu} \xi_{\nu} - \partial_{\nu} \xi_{\mu}\big) = 0 \,.
\label{cov_const}\end{equation}
Thus the Lie derivative with respect to the covariantly constant Killing vector satisfies
\begin{equation}
  \mathcal{L}_{\xi} \mathcal{F}_{\mu_{1}\cdots \mu_{n}}{}^{\nu_{1}\cdots \nu_{m}} = \xi^{\mu} \nabla_{0\mu} \mathcal{F}_{\mu_{1}\cdots \mu_{n}}{}^{\nu_{1}\cdots \nu_{m}} = 0\,,
\label{}\end{equation}
where $\mathcal{F}_{\mu_{1}\cdots \mu_{n}}{}^{\nu_{1}\cdots \nu_{m}}$ is an arbitrary tensor field. The covariantly constant condition constrains the form of the background metric. Let us assume that the Killing vector is of the constant form $\xi = \partial_{z}$, where $z$ is the coordinate along the isometry direction. Since $\xi_{\mu}$ can be locally represented by an exact form, the background metric must satisfy
\begin{equation}
  g_{0}{}_{\mu z} = \partial_{\mu} f(z, x^{i}) \, ,
\label{}\end{equation}
where $f$ is an arbitrary function.

In order to find the single copy, we contract the Killing vector with the KS equations of motion $\hat{\mathcal{R}}_{\hat{\mu}\hat{\nu}}$. Since $\hat{\mathcal{R}}_{\hat{\mu}\hat{\nu}}$ is neither symmetric nor antisymmetric, we obtain two independent sets of equations from \eqref{KS_EOM_KK_reduction}:
\begin{equation}
\begin{aligned}
  \begin{cases}
  \xi^{\mu} \hat{\mathcal{R}}_{\mu\nu} :\quad \nabla_{0}^{\rho}\nabla_{0\rho} \big(\varphi (\xi\cdot l) \bar{l}_{\nu}\big) - \nabla_{0}^{\rho}\nabla_{0\nu} \big(\varphi (\xi\cdot l) \bar{l}_{\mu}\big) = 0 \, ,
  \\
  \xi^{\mu} \hat{\mathcal{R}}_{\mu\alpha} :\quad \nabla_{0}^{\rho}\nabla_{0\rho} \big(\varphi (\xi\cdot l) \bar{j}_{\alpha}\big) = 0 \, ,
  \end{cases}
  \\
  \begin{cases}
  	\xi^{\nu} \hat{\mathcal{R}}_{\mu\nu} :\quad \nabla_{0}^{\rho}\nabla_{0\rho} \big(\varphi (\xi\cdot \bar{l}) l_{\nu}\big) - \nabla_{0}^{\rho}\nabla_{0\nu} \big(\varphi (\xi\cdot \bar{l}) l_{\mu}\big) = 0 \, ,
  	\\
  	\xi^{\nu} \hat{\mathcal{R}}_{\alpha\nu} :\quad \nabla_{0}^{\rho}\nabla_{0\rho} \big(\varphi (\xi\cdot \bar{l}) j_{\alpha}\big) = 0 \, .
  \end{cases}
\end{aligned}\label{contraction_xi_eom}
\end{equation}
Let us define the single copy map as
\begin{equation}
\begin{aligned}
    \mathcal{A}_{\mu} &= \varphi (\xi\cdot \bar{l}) l_{\mu} \,, \qquad X_{\alpha} = \varphi (\xi\cdot \bar{l}) j_{\alpha}\,,
    \\
    \bar{\mathcal{A}}_{\mu} &= \varphi (\xi\cdot l) \bar{l}_{\mu} \,, \qquad \bar{X}_{\alpha} = \varphi (\xi\cdot l) \bar{j}_{\alpha}\,,
\end{aligned}\label{single_copy}
\end{equation}
where $\mathcal{A}$ and $X_{\alpha}$ (similarly $\bar{\mathcal{A}}$ and $\bar{X}_{\alpha}$) are the Maxwell field and $n$ scalar fields, respectively. If we substitute the single copy map into \eqref{contraction_xi_eom}, we obtain
\begin{equation}
\begin{aligned}
  \nabla_0^{\mu} \mathcal{F}_{\mu\nu} = 0\,, \qquad \nabla_{0}^{\rho} \nabla_{0\rho} X_{\alpha} = 0\,,
  \\
  \nabla_{0}^{\mu} \bar{\mathcal{F}}_{\mu\nu} = 0\,, \qquad \nabla_{0}^{\rho} \nabla_{0\rho} \bar{X}_{\alpha} = 0\,,
\end{aligned}\label{}
\end{equation}
where $\mathcal{F}_{\mu\nu}$ and $\bar{\mathcal{F}}_{\mu\nu}$ are the field strengths of $\mathcal{A}$ and $\bar{\mathcal{A}}$, respectively. This shows that the classical double copy for ungauged half-maximal supergravities is given by a pair of Maxwell-scalar theories, the abelian subsector of \eqref{single_copy_action}.

\begin{figure}[bt]
  \includegraphics[scale=0.25]{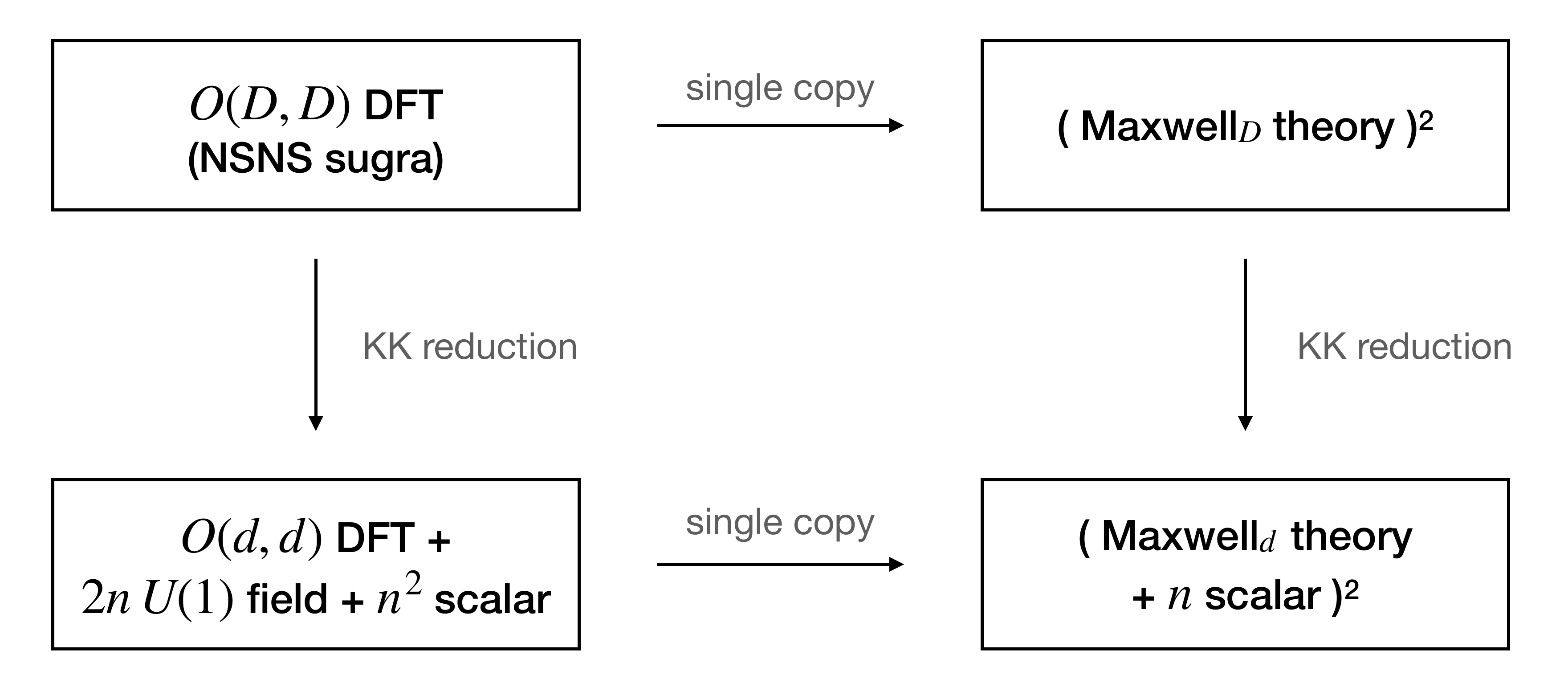}
  \centering
  \caption{A commutative diagram that shows the relations between the single copy and KK reduction. }
\end{figure}

If we substitute the single copy into the KS ansatz for $g$ and $B$ in \eqref{KK_KS_undoubled}, we appear to lose the left-right-mover factorization structure. However, note that the generalised metric before KK reduction is represented by the KK ansatz
\begin{equation}
  \hat{\mathcal{H}}^{\mu\nu} =\mathcal{H}^{\mu\nu} \,, \qquad \hat{\mathcal{H}}^{\mu}{}_{\nu} = \mathcal{H}^{\mu}{}_{\nu} -\frac{1}{2} \mathcal{H}^{\mu\rho} (W W^{t})_{\rho\nu}\,.
\label{}\end{equation}
Using the KS ansatz and the single copy, we can show the following factorization into left and right sectors:
\begin{equation}
\begin{aligned}
  g^{\mu\nu} &\longrightarrow \varphi^{-1} \cA^{(\mu} \barcA^{\nu)}\,,
  \\
  -g^{\mu\rho}\big(B +\tfrac{1}{2} (A\cdot C^{t} + C\cdot A^{t})\big){}_{\rho\nu}  &\longrightarrow  \varphi^{-1} \big(\cA+\cdots\big){}^{\mu} \big(\barcA+\cdots\big){}_{\nu} \, .
\end{aligned}\label{}
\end{equation}
The first equation is not an issue because we can perform the field redefinition $g \to g^{-1}$, after which the new metric perturbation is factorised as we desire. On the other hand, for the $B$-field case, the inner products between $A^{\alpha}$ and $C_{\alpha}$ cannot be removed by a field redefinition of $B$ because these are symmetric. This means we must take into account the $A$ and $C$ contributions in order to observe the manifest double copy for the $B$-field.

\subsection{T-duality rotation}
We now investigate the T-duality transformation of the single copy map. As we have discussed in section \ref{sec:3}, the T-duality group for the $T^{n}$ torus bundle is given by $\mathit{O}(n,n)$. However, here we are assuming that the external space $N_{d}$ admits a globally defined Killing vector $\xi$, and thus it is possible to combine this with the Killing vectors of the internal torus $k^{\alpha}$ such that the total T-duality group is enhanced to $\mathit{O}(n+1,n+1)$. Since our main concern is stationary spacetime, we fix the isometry direction as time. In this setup, the total space is a bundle with an $\mathbb{R}\times T^{n}$ fibre and an $N_{d-1}$ base. Note that $\xi$ is distinct from the Killing vectors for $T^{n}$, $k^{\alpha}$, and commutes with them, $[\xi,k^{\alpha}] = 0$. Thus we can describe the setup as a double KK reduction,  $M_{D} \to N_{d} \to N_{d-1}$, and the corresponding KK ansatz is given by
\begin{equation}
  \hat{g}_{\hat{\mu}\hat{\nu}} = \begin{pmatrix} \check{g}_{ij} + \check{A}_{i}{}^{\check{\alpha}} \check{h}_{\check{\alpha}\check{\beta}} (\check{A}^{t})^{\check{\beta}}{}_{\nu} & \check{A}_{i}{}^{\check{\gamma}} \check{h}_{\check{\gamma}\check{\beta}} \\ \check{h}_{\check{\alpha}\check{\gamma}}  (\check{A}^{t})^{\check\gamma}{}_{j} & \check{h}_{\check{\alpha}\check{\beta}}\end{pmatrix} \, ,
\label{}\end{equation}
where
\begin{equation}
\begin{aligned}
  \check{g}_{ij} &= g_{ij} \,,
  \qquad
  \check{A}_{i}{}^{\check{\alpha}} &= \begin{pmatrix} \frac{g_{ti}}{g_{tt}} \\ A_{i}{}^{\alpha}-\frac{g_{ti}}{g_{tt}} A_{t}{}^{\alpha} \end{pmatrix}\,,
  \qquad
  \check{h}_{\check{\alpha}\check{\beta}} = \begin{pmatrix} g_{tt} + A_{t}{}^{\alpha} h_{\alpha\beta} (A^{T})^{\beta}{}_{t} & A_{t}{}^{\alpha} h_{\alpha\beta}\\ h_{\alpha\beta} (A^{T})^{\beta}{}_{t} & h_{\alpha\beta} \end{pmatrix}\,.
\end{aligned}\label{}
\end{equation}

Let us consider the KS ansatz and assume that the background is an $\mathbb{R}\times T^{n}$ bundle. We introduce coordinates for the base such that the timelike Killing vector is a constant, $\xi = \frac{\partial}{\partial t}$, and we append the time coordinate to the torus coordinates $y^{\alpha}$ as $\check{y}^{\check{\alpha}}= (t,y^{\alpha})$. Then the remaining coordinates for $N_{d-1}$ are given by $\{x^{i}\}$. We denote the background bundle connection as $\check{A}_{0}^{\check{\alpha}}$ and decompose the null vectors $\hat{l}$ and $\bar{l}$ as
\begin{equation}
  \hat{l}_{\hat{\mu}} = \begin{pmatrix} l_{i}+\check{A}_{0i}{}^{\check{\alpha}} \check{j}_{\check{\alpha}} \\ \check{j}_{\check{\alpha}} \end{pmatrix}\,,
  \qquad
  \hat{\bar{l}}_{\hat{\mu}} = \begin{pmatrix} \bar{l}_{i}+\check{A}_{0i}{}^{\check{\alpha}} \check{j}_{\check{\alpha}} \\ \check{j}_{\check{\alpha}} \end{pmatrix} \,,
\label{}\end{equation}
where the fibre components $\check{j}_{\check{\alpha}}$ and $\check{\bar{j}}_{\check{\alpha}}$ are given by
\begin{equation}
  \check{j}_{\check{\alpha}} = \begin{pmatrix} l_{t} \\ j_{\alpha} \end{pmatrix}\,,
  \qquad
  \check{\bar{j}}_{\check{\alpha}} = \begin{pmatrix} \bar{l}_{t} \\ \bar{j}_{\alpha} \end{pmatrix}\,.
\label{}\end{equation}

The $\mathit{O}(n+1,n+1)$ rotation affects only $\check{j}_{\check{\alpha}}$ and $\check{\bar{j}}_{\check{\alpha}}$,
\begin{equation}
\begin{aligned}
  \check{j}'^{\check\alpha} &= \check{a}^{\check{\alpha}}{}_{\check{\beta}} \check{j}^{\check{\beta}} + \check{b}^{\check{\alpha}\check{\beta}} E_{0\check{\beta}\check{\gamma}}j^{\check{\gamma}} \,,
  \\
  \check{\bar{j}}'^{\check{\alpha}} &= \check{a}^{\check{\alpha}}{}_{\check{\beta}} \bar{j}^{\check{\beta}} - \check{b}^{\check{\alpha}\check{\beta}} (E^{T}_{0})_{\check{\beta}\check{\gamma}} \check{\bar{j}}^{\check{\gamma}}\,.
\end{aligned}\label{double_boosts_jjbar}
\end{equation}
Strictly speaking, the full $\mathit{O}(n+1,n+1)$ rotation cannot be T-duality because time is not a compact direction. Thus it is merely a solution-generating transformation rather than a duality rotation. We will consider the (Lorentz analogue of the) maximal compact subgroup of $\mathit{O}(n+1,n+1)$, $\mathit{O}(1,n)^{2}$, in order to preserve the asymptotic structure. Among the $\mathit{O}(1,n)$ group, we will focus only on the Lorentz boost part $\mathit{SO}(1,1)$ because the $T^{n}$ rotation part $\mathit{SO}(n)$ leaves $l_{\mu}$ and $\bar{l}_{\mu}$ unchanged. This implies that $\mathit{SO}(n)$ does not transform $g_{\mu\nu}$ and $B_{\mu\nu}$ and acts only on the $\mathit{U}(1)$ gauge fields and moduli scalars: in order to obtain a new metric and $B$-field, we must consider the Lorentz boost parts of the transformation.

As a simple example, we will illustrate the case where the internal torus is a circle, corresponding to $n=1$.
The $\mathit{SO}(1,1)^{2}$ transformation can be embedded into the $\mathit{O}(2,2)$ duality group as
\begin{equation}
  \mathcal{O}_{\check{A}}{}^{\check{B}} = \begin{pmatrix} a^{\check{\alpha}}{}_{\check{\beta}} & b^{\check{\alpha}\check{\beta}} \\ c_{\check{\alpha}\check{\beta}} & d_{\check{\alpha}}{}^{\check{\beta}} \end{pmatrix}
  = \frac{1}{2} \begin{pmatrix} (\check{o}_{1}+\check{o}_{2})^{\check{\alpha}}{}_{\check{\beta}} & (\check{o}_{1}-\check{o}_{2})^{\check{\alpha}}{}_{\check{\gamma}} \check{\eta}^{\check{\gamma}\check{\beta}}
  \\ \check{\eta}_{\check{\alpha}\check{\gamma}} (\check{o}_{1}-\check{o}_{2})^{\check{\gamma}}{}_{\check{\beta}} & \check{\eta}_{\check{\alpha}\check{\gamma}} (\check{o}_{1} + \check{o}_{2})^{\check{\gamma}}{}_{\check{\delta}} \check{\eta}{}^{\check{\delta}\check{\beta}} \end{pmatrix}\,,
\label{}\end{equation}
where
\begin{equation}
  \check{o}_{1} = \begin{pmatrix} \mathbf{1} & 0 & 0 \\ 0 & \cosh{\alpha} & \sinh{\alpha} \\ 0 & \sinh{\alpha} & \cosh{\alpha} \end{pmatrix} \,,
  \qquad
  \check{o}_{2} = \begin{pmatrix} \mathbf{1} & 0 & 0 \\ 0 & \cosh{\beta} & \sinh{\beta} \\ 0 & \sinh{\beta} & \cosh{\beta}\end{pmatrix} \,,
  \qquad
  \check{\eta} = \begin{pmatrix} \mathbf{1} & 0 & 0 \\ 0 & -1 & 0 \\ 0 & 0 & 1 \end{pmatrix} \,.
\label{}\end{equation}
The double Lorentz boosts transform the single copy as
\begin{equation}
\begin{aligned}
  \begin{pmatrix} \mathcal{A}'_{t} \\ \mathcal{A}'_{i} \end{pmatrix} = \begin{pmatrix} \cosh\alpha \mathcal{A} + \sinh\alpha X \end{pmatrix} \, .
\end{aligned}\label{}
\end{equation}
As a special case, if the background metric is trivial, $g_{0} = \eta$, the double Lorentz boost transformation of the single copy reduces to
\begin{equation}
\begin{aligned}
  \begin{pmatrix} \mathcal{A}'_{t} \\ \mathcal{A}'_{i} \\ X_{z} \end{pmatrix} = \begin{pmatrix} \cosh\alpha \mathcal{A}_{t} + \sinh\alpha X_{z} \\ \mathcal{A}_{i} \\ \sinh \alpha \mathcal{A}_{t} + \cosh\alpha X_{z}\end{pmatrix} \,,
  \qquad
  \begin{pmatrix} \bar{\mathcal{A}}'_{t} \\ \bar{\mathcal{A}}'_{i} \\ \bar{X}_{z} \end{pmatrix} = \begin{pmatrix} \cosh\beta \bar{\mathcal{A}}_{t} + \sinh\beta \bar{X}_{z} \\ \bar{\mathcal{A}}_{i} \\\sinh\beta \bar{\mathcal{A}}_{t} + \cosh\beta \bar{X}_{z}\end{pmatrix} \, .
\end{aligned}\label{double_boosts_single_copy}
\end{equation}
Thus in the four-dimensional case, the transformation mixes the electric fields and scalar fields, but leaves the magnetic fields invariant.

\section{Examples} \label{sec:5}

\subsection{Buscher transformation for the Kerr BH} \label{sec:5.1}
The Kerr BH is a stationary axi-symmetric solution of the vacuum Einstein equations. It admits two Killing vectors: $\partial_{t}$ and $\partial_{\phi}$.   Let us apply the Buscher rule along the time direction. The KS ansatz for the Kerr BH is given by
\begin{equation}
  \mathrm{d}s^{2} = \mathrm{d}s_{0}^{2} - \kappa\varphi l l \, ,
\label{}\end{equation}
where $\kappa\varphi = -\frac{2 m r^{3}}{r^{4}+a^{2} z^{2}}$ and
\begin{equation}
 l = \mathrm{d} t+\frac{r(x \mathrm{~d} x+y \mathrm{~d} y)}{a^{2}+r^{2}}+\frac{a(y \mathrm{~d} x-x \mathrm{~d} y)}{a^{2}+r^{2}}+\frac{z}{r} \mathrm{~d} z \,.
\label{KerrBH_nullvector}\end{equation}
Here $r$ is a function 	satisfying
\begin{equation}
  x^{2}+y^{2}+z^{2}=r^{2}+a^{2}\left(1-\frac{z^{2}}{r^{2}}\right)\,.
\label{}\end{equation}
Now we apply the Buscher transformation for the KS ansatz \eqref{buscher_rule_KS} because the background is trivial. Since the null vectors are identical, $\bar{l}=l$, there is one independent transformation among the three cases (the third transformation gives a solution identical to the Kerr BH),
\begin{equation}
\begin{aligned}
  \mathrm{I} &~ :  l' = \begin{pmatrix} 1 \,, & \dfrac{rx+ay}{r^2+a^2}\,, & \dfrac{ry-ax}{r^2+a^2}\,,  & \dfrac{z}{r} \end{pmatrix} \,, \qquad \bar{l}' = \begin{pmatrix} -1 \,, & \dfrac{rx+ay}{r^2+a^2}\,, & \dfrac{ry-ax}{r^2+a^2}\,,  & \dfrac{z}{r} \end{pmatrix} \, .
\end{aligned}\label{}
\end{equation}
If we substitute this result into the KS ansatz, we obtain a new solution with a nonvanishing Kalb-Ramond field and nontrivial dilaton. We may represent the solution in spheroidal coordinates via a coordinate transformation,
\begin{equation}
\begin{aligned}
  \mathrm{d}s^2 &= \frac{r^2+ a^2 \cos^2\theta}{r^2 - 2mr + a^2 \cos^2\theta} \Bigg[ \ -\mathrm{d}t^2  + \frac{r^2 -2mr + a^2 \cos^2\theta}{r^2 - 2mr+a^2 } \mathrm{d} r^2 +(r^2 - 2mr+a^2\cos^{2}\theta)\mathrm{d}\theta^{2}
  \\
  &\qquad\qquad\qquad\qquad\qquad\quad +  (r^2 - 2mr+a^2) \sin^2\theta d\phi^2 ~\Bigg]\,,
  \\
  B_{t \phi} &= \frac{2 mar \sin^{2}\theta}{r^{2}-2 m r+a^{2} \cos^2\theta}\,,
  \qquad
  e^{2\phi} = \frac{r^{2}+a^{2} \cos ^{2}\theta}{r^{2}-2 m r+a^{2} \cos ^{2}\theta}\,.
\end{aligned}\label{}
\end{equation}
This is written in string frame, and we can convert to Einstein frame via the Weyl rescaling $\mathrm{d}s^{2}_{E} = e^{-2\phi} \mathrm{d}s^{2}$. One can check that this solution is asymptotically flat. Since $g^{E}_{tt} = 1$ and $g^{E}_{t\phi}=0$, it is apparent that there is no horizon and no angular momentum. Thus this solution has a naked singularity.

The single copy for this solution is of the same form as the Kerr BH except for the signature of the electric charges.
\begin{equation}
  Q = Q_{\mathrm{Kerr}}\,, \qquad \bar{Q} = -Q_{\mathrm{Kerr}} \,,
\label{}\end{equation}
where $Q$ and $\bar{Q}$ are charges for $\mathcal{A}_{0}$ and $\bar{\mathcal{A}}_{0}$, respectively, and $Q_{\mathrm{Kerr}}$ is the single copy charge for the Kerr BH obtained in \cite{Monteiro:2014cda}.
It is interesting that such a simple modification in the single copy generates a huge difference in the supergravity solution.

\subsection{Sen's BH in heterotic supergravity}
In \cite{Sen:1994eb} Sen constructed the most general electrically charged rotating black hole solutions of heterotic supergravity compactified on $T^{6}$. This solution is obtained via a T-duality transformation of the Kerr BH solution of the vacuum Einstein equations. In order to apply the T-duality rotation, the Kerr BH should first be embedded into heterotic supergravity by turning off all fields other than the metric. Next, we apply the $\mathit{O}(6,22)$ T-duality rotation, which is associated with $T^{6}$ compactification of the heterotic string. It is important to note that this duality rotation acts only on the internal torus components, thus cannot be used to generate new solutions.  In order to nevertheless generate nontrivial solutions, Sen's strategy was to incorporate the time direction into the $T^{6}$ isometry. Thus the duality group is enhanced to $\mathit{O}(7,23)$ and can generate new solutions via mixing of external and internal components of the metric. The rotation also generates $B$-field, dilaton and electric fields, and the rotation parameters are identified with their charges.

We now construct the KS ansatz for Sen's BH following the same procedure. Starting from the KS ansatz for the Kerr BH, we apply T-duality and perform the dimensional reduction. For simplicity, we consider KK reduction of $\mathit{O}(5,5)$ DFT without vector multiplets, rather than ten-dimensional heterotic supergravity on $T^{6}$. Although this is not the most general setup, the procedure can easily be generalised to other cases. Using the KS ansatz, we may identify the single copy of Sen's BH. As we have seen in section \ref{sec:4}, the duality transformation may be enhanced to $\mathit{O}(2,2)$ by including the time direction in the isometry together with the circle direction. Among the elements of $\mathit{O}(2,2)$, we consider only the Lorentzian version of the maximal compact subgroup, $\mathit{SO}(1,1) \times \mathit{SO}(1,1)$, or double Lorentz boosts, in order to preserve the asymptotically flat structure of the solution. Strictly speaking, the Lorentz boosts are not T-duality transformations because the time direction is non-compact. However, we can use them as a solution generating technique rather than duality transformations. Thus the transformed solutions are not related to the seed solution, in this case the Kerr BH, by string dualities.

Let us now describe the procedure in more detail.  First of all, we wish to embed the Kerr BH solution into the $\mathit{O}(5,5)$ DFT KS ansatz. The corresponding $\mathit{O}(5,5)$ DFT null vectors are
\begin{equation}
  \hat{l}_{\hat{\mu}} = \begin{pmatrix} l_{\mu} \\ 0\end{pmatrix} \,,
  \qquad
  \hat{\bar{l}}_{\hat{\mu}} = \begin{pmatrix} \bar{l}_{\mu} \\ 0\end{pmatrix}\,,
\label{}\end{equation}
where $l_{\mu}$ and $\bar{l}_{\mu}$ are identical and defined in \eqref{KerrBH_nullvector}.

We then apply the $\mathit{SO}(1,1)^{2}\subset \mathit{O}(2,2)$ rotation to the DFT KS ansatz. The double Lorentz boosts $o_{1,2}$ along the internal circle direction are represented by
\begin{align}
  o_{1} = \begin{pmatrix}
  \cosh\alpha & 0 & \sinh\alpha \\ 0 & \mathbf{1}_{3} & 0\\ \sinh\alpha & 0 & \cosh\alpha
  \end{pmatrix}\,,
  \qquad
  o_{2} = \begin{pmatrix}
  \cosh\beta & 0 & \sinh\beta \\ 0 & \mathbf{1}_{3} & 0\\ \sinh\beta & 0 & \cosh\beta
  \end{pmatrix}\,.
\end{align}
The corresponding $\mathit{O}(2,2)$ matrix is
\begin{equation}
  \mathcal{O}_{\hat{M}}{}^{\hat{N}} = \begin{pmatrix} a^{\hat{\mu}}{}_{\hat{\nu}} & b^{\hat{\mu}\hat{\nu}} \\ c_{\hat{\mu}\hat{\nu}} & d_{\hat{\mu}}{}^{\hat{\nu}} \end{pmatrix}
  = \frac{1}{4} \begin{pmatrix} (o_{1}+o_{2})^{\hat{\mu}}{}_{\hat{\nu}} & (o_{1}-o_{2})^{\hat{\mu}}{}_{\hat{\rho}}\hat{\eta}^{\hat{\rho}\hat{\nu}} \\ \hat{\eta}_{\hat{\mu}\hat{\rho}} (o_{1}-o_{2})^{\hat{\rho}}{}_{\hat{\nu}} & \hat{\eta}_{\hat{\mu}\hat{\rho}} (o_{1}-o_{2})^{\hat{\rho}}{}_{\hat{\sigma}} \hat{\eta}^{\hat{\sigma}\hat{\nu}} \end{pmatrix}\,,
\label{}\end{equation}
where $\hat{\eta}$ is the five-dimensional flat metric, $\hat{\eta} = \text{diag}(-1,1,1,1,1)$. Since the background $B$-field vanishes, one can show that the background metric is invariant under the double Lorentz boost,
\begin{equation}
  a \hat{\eta}^{-1} a^{t} + b \hat{\eta} b^{t}= \hat{\eta}^{-1}\,.
\label{}\end{equation}
From \eqref{o12} and \eqref{o34}, the left and right null vectors for a trivial background metric transform linearly, $\hat{l}' = o_{1} \hat{l}$ and $\hat{\bar{l}}' = o_{2} \hat{\bar{l}}$, and their components are given by
\begin{equation}
  \hat{l}'_{\hat{\mu}} = \begin{pmatrix} \cosh\alpha \\ l_{i} \\ \sinh\alpha \end{pmatrix}\,,
  \qquad
  \hat{\bar{l}}'_{\hat{\mu}} = \begin{pmatrix} \cosh\beta \\ \bar{l}_{i} \\ \sinh\beta \end{pmatrix}\,,
\label{}\end{equation}
where $l_{i}$ and $\bar{l}_{i}$ are the spatial components of \eqref{KerrBH_nullvector}, and $\varphi$ and $\varphi'$ defined in \eqref{varphip} are
\begin{equation}
\begin{aligned}
  \varphi=-\frac{2 m r^{3}}{r^{4}+a^{2} z^{2}} \, , \qquad & \varphi' = -\frac{mr}{\Sigma - mr \left(1-\cosh\alpha \cosh\beta \right)}\,,
\end{aligned}
\end{equation}
where $\Sigma = r^2+ \frac{a^2 z^{2}}{r^{2}} $. We can thus read off the supergravity fields from the KS ansatz in \eqref{KK_KS_undoubled}:
\begin{equation}
\begin{aligned}
  \mathrm{d}s^{2} &= \eta_{\mu\nu} \mathrm{d}x^{\mu} \mathrm{d}x^{\nu} +\frac{2mr\big(\Sigma - mr(1-\cosh\alpha\cosh\beta)\big)}{\Delta} l' \bar{l}' - \frac{m^{2}r^{2}}{\Delta}(\sinh^{2}\beta\ l'l' + \sinh^{2}\alpha\ \bar{l}'\bar{l}') \, ,
  \\
  B &= -\frac{mr\big(\Sigma - mr(1-\cosh\alpha\cosh\beta)\big)}{\Delta}\left(\cosh\alpha - \cosh\beta\right) \mathrm{d}t \wedge l' \, ,
  \\
  A &=  -\frac{mr}{\Sigma -mr(1-\cosh(\alpha-\beta))} (-\sinh\beta l' + \sinh\alpha \bar{l}') \, ,
  \\
  C &=  -\frac{mr}{\Sigma -mr(1-\cosh(\alpha+\beta))} (\sinh\beta l' + \sinh\alpha \bar{l}') \, ,
  \\
  h &= 1 - \frac{2mr \sinh\alpha \sinh\beta}{\Sigma -mr\left(1-\cosh(\alpha + \beta) \right)}\,,
\end{aligned}
\end{equation}
where
\begin{equation}
\begin{aligned}
  \Delta &= \Sigma^{2} -2mr \Sigma (1-\cosh \alpha \cosh \beta) + m^{2}r^{2} (\cosh\alpha-\cosh\beta)^{2}
  \\
  & = \big(\Sigma-mr(1-\cosh(\alpha+\beta))\big)\big(\Sigma-mr(1-\cosh(\alpha-\beta))\big)\,.
\end{aligned}\label{}
\end{equation}

\subsubsection{Coordinate transformation}
To compare the above results to Sen's BH \cite{Sen:1994eb}, we consider a coordinate transformation from KS coordinates to the Boyer--Lindquist coordinates
\begin{equation}
\begin{aligned}
  \mathrm{d}t &\longrightarrow \mathrm{d}t +  \frac{mr( \cosh\alpha + \cosh\beta)}{r^2 + a^2 -2mr} \mathrm{d}r \, ,
  \\
  \mathrm{d}x &\longrightarrow \sin\theta\left( \cos\bar{\phi}    - \frac{a \left( r \sin\bar{\phi} + a \cos\bar{\phi}\right)}{r^2+a^2 -2mr}
    \right)\mathrm{d}r
  \\
  &\qquad + \cos\theta \left( r \cos\bar{\phi} - a \sin\bar{\phi}\right) \mathrm{d}\theta - \left( r\sin\bar{\phi} +a\cos\bar{\phi} \right) \sin\theta \mathrm{d}\phi  \, ,
  \\
  \mathrm{d}y &\longrightarrow \sin\theta \left(  \sin\bar{\phi} + \frac{a  \left( r \cos\bar{\phi} - a \sin\bar{\phi}\right)}{r^2+ a^2 -2mr} \right) \mathrm{d}r
  \\
    & \qquad + \cos\theta \left( r \sin\bar{\phi} + a \cos\bar{\phi} \right) \mathrm{d}\theta + \sin\theta \left(r\cos\bar{\phi} - a \sin\bar{\phi} \right) \mathrm{d}{\phi} \, ,
    \\
  \mathrm{d}z &\longrightarrow \cos\theta \mathrm{d}r -r\sin\theta \mathrm{d}\theta \, ,
\end{aligned}\label{}
\end{equation}
where $\bar{\phi} = \phi + \int\frac{a}{r^2+a^2 -2mr} \mathrm{d}r$. Applying this coordinate transformation, the fields match Sen's solution up to gauge transformations of the form fields, $A$, $B$ and $C$:
\begin{equation}
\begin{aligned}
  \mathrm{d}s^2 &=  \Sigma\Big(- \frac{( \Sigma - 2mr)}{\Delta}  \mathrm{d} t^2 - \frac{ 2m r a \sin^2\theta }{\Delta} (\cosh\alpha + \cosh\beta) \mathrm{d}t \mathrm{d}\phi
  + \frac{\mathrm{d}r^2 }{r^2 + a^2 - 2mr} + \mathrm{d}\theta^2  \\
  &\qquad~~ + \frac{\sin^{2}\theta}{\Delta}\big(\Delta+ a^2 \sin^2\theta \left( \Sigma + 2 m r \cosh\alpha \cosh\beta \right)\big) \mathrm{d}\phi^2 \Big) \, ,
  \\
  B &= - \Delta^{-1} m r a \sin^2\theta \left( \cosh\alpha - \cosh\beta \right)\big( \Sigma -mr \left(1- \cosh\alpha \cosh\beta \right)\big) \mathrm{d}t \wedge \mathrm{d}\phi \, ,
  \\
  A &= \frac{m r }{\Sigma -m r\left(1-\cosh \left(\alpha-\beta\right)\right)} \big(-\sinh \left(\alpha-\beta\right)\mathrm{d}t + a \sin^{2}\theta\left(\sinh \alpha -\sinh \beta\right) \mathrm{d}\phi\big) \, ,
  \\
  C &=\frac{m r }{\Sigma -m r\left(1-\cosh \left(\alpha+\beta\right)\right)} \big(-\sinh \left(\alpha+\beta\right)\mathrm{d}t + a \sin^{2}\theta\left(\sinh \alpha +\sinh \beta\right) \mathrm{d}\phi\big) \, ,
  \\
  h &= 1 - \frac{2mr \sinh\alpha \sinh\beta}{\Sigma -mr\left(1-\cosh(\alpha + \beta) \right)}\,,
  \\
  \phi &= \frac{1}{2} \ln \frac{\left(\rho^{2}+a^{2} \cos ^{2} \theta\right)^{2}}{\Delta} \, .
\end{aligned}\label{}
\end{equation}

\subsubsection{Classical double copy}
Sen's BH has a time-like Killing vector which satisfies the covariantly-constant condition \eqref{cov_const},
\begin{equation}
  \xi = \frac{\partial}{\partial t} \, .
\label{}\end{equation}
The inner products between the null vectors and the Killing vector are
\begin{equation}
  \xi \cdot l = \cosh \alpha\,, \qquad \xi \cdot \bar{l} = \cosh\beta\,.
\label{}\end{equation}
From the single copy map \eqref{single_copy}, we obtain the corresponding pair of Maxwell and scalar fields:
\begin{equation}
\begin{aligned}
  A &= -\frac{2mr^{3}\cosh\beta}{r^{4}+a^{2}z^{2}}\begin{pmatrix} \cosh\alpha \\ \frac{rx+ay}{r^{2}+a^{2}} \\ \frac{ry-ax}{r^{2}+a^{2}} \\ \frac{z}{r} \end{pmatrix} \,,
  &\qquad
  \bar{A} &= -\frac{2mr^{3}\cosh\alpha}{r^{4}+a^{2}z^{2}}\begin{pmatrix} \cosh\beta \\ \frac{rx+ay}{r^{2}+a^{2}} \\ \frac{ry-ax}{r^{2}+a^{2}} \\ \frac{z}{r} \end{pmatrix} \,, \
  \\
  X &= -\frac{2mr^{3}\sinh\alpha\cosh\beta}{r^{4}+a^{2}z^{2}} \,,
  &\qquad
  \bar{X} &= -\frac{2mr^{3}\cosh\alpha\sinh\beta}{r^{4}+a^{2}z^{2}} \,.
\end{aligned}\label{}
\end{equation}

The $\mathit{SO}(1,1)^{2}$ rotation maps the electromagnetic charges and currents of the single copy of the Kerr BH \cite{Monteiro:2014cda} into
\begin{equation}
\begin{aligned}
  Q &= \cosh\alpha \cosh\beta Q_{0} \, , &\qquad \bar{Q} &= \cosh\alpha \cosh\beta Q_{0}\,,
  \\
  I &= \cosh\beta I_{0} \, , &\qquad \bar{I} &= \cosh\alpha I_{0}\,,
  \\
  q &= 2m \sin\alpha\cosh\beta \, , &\qquad \bar{q} &= 2	m \cosh\alpha\sin\beta \, ,
\end{aligned}\label{}
\end{equation}
where $Q_{0}$ and $I_{0}$ are the charges and currents of the single copy of the Kerr BH solution. If we set the boost parameters as $\alpha=\beta=0$, the solution reduces to the Kerr BH case.

\subsection{Chiral null model}
The chiral null model (CNM) describes a class of half-maximal supergravity solutions, which corresponds to a class of exact closed string backgrounds associated with a chiral null current \cite{Horowitz:1994rf}. 
It covers various types of solutions, including a string theory generalisation of the Taub--NUT solution and the rotating BH. For the 5-dimensional case, we may decompose the coordinates as $\hat{x}^{\hat{\mu}} = (v,x^{i},u)$, where $u$ and $v$ correspond to null directions.
The metric, B-field and dilaton take the form\footnote{The CNM admits $u$-dependence in general, but we will neglect the $u$-dependence for consistent dimensional reduction.}
\begin{align}
	\rd\hs^{2} &= F(x^{i})\rd u\big(\rd v + K(x^{i})\rd u + 2V_{i}(x^{i})\rd x^{i}\big) + \delta_{ij}\rd x^{i}\rd x^{j} \, , \label{CNMmetric} \\
	\hB_{uv} &= -\frac{1}{2}F(x^{i}) \, , \qquad \hB_{ui} = -F(x^{i})V_{i}(x^{i}) \, , \\
	\hphi &= \frac{1}{2}\ln F(x^{i}) \, ,
\end{align}
where $F$, $K$ and $V_{i}$ satisfy
\begin{equation}
  \partial^{i} \partial_{i} F^{-1} = 0\,, \qquad \partial^{i} \partial_{i} K = 0\,, \qquad \partial^{i} \big(\partial_{i} V_{j} - \partial_{j} V_{i}\big) = 0\,. \label{CNM_eom}
\end{equation}
We may write this in KS form via a coordinate transformation
\begin{equation}
	\tv = v + \Psi(x^{i}) \, , \label{vtilde}
\end{equation}
where in the following, $\Psi(x^{i})$ is chosen to be independent of $u$.  This ensures that the bundle structure is preserved under the KS decomposition.  Defining
\begin{equation}
	\tV_{i} \equiv V_{i} - \frac{1}{2}\p_{i}\Psi \, ,
\end{equation}
the ansatz for the metric becomes
\begin{equation}
	\rd s^{2} = \rd u \rd\tv + \delta_{ij}\rd x^{i}\rd x^{j} + (F-1)\rd u\left(\rd\tv + \frac{KF}{(F-1)}\rd u + \frac{2F}{(F-1)}\tV_{i}\rd x^{i}\right) \, ,
\end{equation}
from which we can read off the scalar function and vectors
\begin{equation}
	\kappa\varphi = F^{-1} - 1 \, , \qquad \hl = \rd u \, , \qquad \hbarl = \rd\tv + \frac{F}{(F-1)}\left(K\rd u + 2\tV_{i}\rd x^{i}\right) \, .
\end{equation}

We can also identify the background metric and its inverse,
\begin{equation}
	\htg_{\hmu\hnu} = \begin{pmatrix} 0 & 0 & \frac{1}{2} \\ 0 & \delta_{ij} & 0 \\ \frac{1}{2} & 0 & 0 \end{pmatrix} \, , \qquad \htg^{\hmu\hnu} = \begin{pmatrix} 0 & 0 & 2 \\ 0 & \delta^{ij} & 0 \\ 2 & 0 & 0 \end{pmatrix} \, . \label{gginvCNM}
\end{equation}
Imposing the null conditions $\hl^2 = \htg^{\hmu\hnu}\hl_{\hmu}\hl_{\hnu} = 0$ and $\hbarl^2 = \htg^{\hmu\hnu}\hbarl_{\hmu}\hbarl_{\hnu} = 0$, we find the constraint
\begin{equation}
	K = - \frac{F}{(F-1)}\tV_{i}\tV^{i} \, , \label{nulluindep}
\end{equation}
giving
\begin{equation}
	\hl = \rd u \, , \qquad \hbarl = \rd\tv - \frac{F^{2}}{(F-1)^{2}}\tV_{i}\tV^{i}\rd u + \frac{2F}{(F-1)}\tV_{i}\rd x^{i} \, .
\end{equation}
Note that the null constraint \eqref{nulluindep} involves only the magnitude of $\tV_{i}$, which we are free to choose since the equations of motion \eqref{CNM_eom} are invariant under constant rescalings of $\tV_{i}$.

Having obtained the KS ansatz, we may transform back to $(u,v)$ coordinates.
%
%
%
For $u$-independent solutions, i.e. solutions satisfying
$\p_{u}\Psi = 0$,
the background metric reduces to
\begin{equation}
	\mathrm{d}\hts^{2} = \mathrm{d}u \mathrm{d}v + \partial_{i}\Psi\mathrm{d}u \mathrm{d}x^{i}+ \delta_{ij} \mathrm{d}x^{i} \mathrm{d}x^{i}\,.
	\label{CNMbackground-uv}
\end{equation}
However, standard compactification of \eqref{CNMbackground-uv} along the $u$-direction is impossible because the $\tg_{uu}$ component vanishes, so the reduced background metric becomes singular.  Thus we need to take a coordinate transformation that shifts $t = v - Yu$ for some $Y\in\mathbb{R}$,
\begin{equation}
	\mathrm{d}\hts^{2} = \mathrm{d}u \mathrm{d}t  + Y\mathrm{d}u\mathrm{d}u + \partial_{i}\Psi\mathrm{d}u \mathrm{d}x^{i} + \delta_{ij} \mathrm{d}x^{i} \mathrm{d}x^{j}\,.
	\label{CNMgbackground-ut}\end{equation}
The background metric is flat and the Riemann tensor vanishes. The corresponding KK ansatz is given by
\begin{equation}
	\begin{aligned}
		\mathrm{d}\ts^{2} &= \delta_{ij} \mathrm{d}x^{i}\mathrm{d}x^{j} -\frac{1}{4Y} \big(\mathrm{d}t+\partial_{i}\Psi \mathrm{d}x^{i}\big)^{2}\,, \qquad \th_{uu} = Y\,,
		\\
		\tA &= \frac{1}{2Y} \big(\mathrm{d}t + \partial_{i}\Psi \mathrm{d}x^{i}\big) \,.
	\end{aligned}\label{CNMreduced}
\end{equation}
Here we have kept the explicit $Y$-dependence, which illustrates that the reduced background metric and gauge field become singular as $Y = 0$.\footnote{Note that the DFT framework includes also non-Riemannian/non-Lorentzian solutions, of which one type is non-relativistic gravities.  It would be interesting to explore the possible connection to the $Y\rightarrow 0$ limit of \eqref{CNMreduced}.}
As a conventional choice we may take $Y = 1$, in which case $t$ is the usual time coordinate.  The reduced background metric and gauge fields are still flat.

\subsubsection{$\mathit{O}(2,2)$ transformation} \label{sec:O22CNM}
Let us consider an $\mathit{O}(2,2)$ rotation of the CNM. Here we focus on the $\mathit{O}(1,1)^{2}$ subgroup only. The transformation matrix $\mathcal{O}$ is given by
\begin{equation}
	\mathcal{O}_{\hat{M}}{}^{\hat{N}} = \begin{pmatrix} a^{\hat{\mu}}{}_{\hat{\nu}} & b^{\hat{\mu}\hat{\nu}} \\ c_{\hat{\mu}\hat{\nu}} & d_{\hat{\mu}}{}^{\hat{\nu}}\end{pmatrix}
	= \frac{1}{2} \begin{pmatrix} (o_{1}+o_{2})^{\hat{\mu}}{}_{\hat{\nu}} & (o_{1}-o_{2})^{\hat{\mu}}{}_{\hat{\rho}}\hat{\eta}^{\hat{\rho}\hat{\nu}} \\ \hat{\eta}_{\hat{\mu}\hat{\rho}} (o_{1}-o_{2})^{\hat{\rho}}{}_{\hat{\nu}} & \hat{\eta}_{\hat{\mu}\hat{\rho}} (o_{1}+o_{2})^{\hat{\rho}}{}_{\hat{\sigma}} \hat{\eta}^{\hat{\sigma}\hat{\nu}} \end{pmatrix}\,,
	\label{CNM_Oabcd}\end{equation}
where $o_{1}$ and $o_{2}$ are $\mathit{SO}(1,1)$ rotations which are represented by
\begin{equation}
	o_1 = \begin{pmatrix} e^{-\alpha} & 0 & 0\\ 0 & \mathbf{1} & 0 \\ 0 & 0 & e^{\alpha} \end{pmatrix}\,, \qquad
	o_2 = \begin{pmatrix} e^{-\beta} & 0 & 0\\ 0 & \mathbf{1} & 0 \\ 0 & 0 & e^{\beta} \end{pmatrix}\,.
	\label{}\end{equation}
We apply the transformation in $(u,t)$ coordinates, in which the background metric \eqref{CNMgbackground-ut} and its inverse are
\begin{equation}
	\htg_{\hmu\hnu} = \begin{pmatrix} 0 & 0 & \frac{1}{2} \\ 0 & \delta_{ij} & \frac{1}{2}\p_{i}\Psi \\ \frac{1}{2} & \frac{1}{2}\p_{j}\Psi & Y \end{pmatrix}\,, \qquad
	\htg^{\hmu\hnu} = \begin{pmatrix} \p_{k}\Psi\p^{k}\Psi - 4Y & -\p^{j}\Psi & 2\\ -\p^{i}\Psi & \delta^{ij} & 0 \\ 2 & 0 & 0 \end{pmatrix}\,, \label{gginvCNMut}
\end{equation}
while the null vectors are written as
\begin{equation}
	\hl = \rd u \, , \qquad \hbarl = \rd t + \left[\frac{2F}{(F-1)}\tV_{i} + \p_{i}\Psi\right]\rd x^{i} + \left[Y - \frac{F K}{(F-1)}\right]\rd u \, . \label{CNMlbarltu}
\end{equation}

Since $\hat{B}_{0} = 0$, after the $\mathit{O}(1,1)^{2}$ transformation the background inverse metric is given by \eqref{Onnginv},
\begin{equation}
	\hat{g}_{0}'^{-1} = a \hat{g}_{0}^{-1} a^{t} + b \hat{g}_{0} b^{t}\,,
	\label{}\end{equation}
from which we obtain
\begin{equation}
	\begin{aligned}
		\mathrm{d}\hs'_{0}{}^{2} = \mathrm{d}u \mathrm{d}t + e^{-\alpha-\beta}Y \mathrm{d}u\mathrm{d}u +\frac{1}{4} \big(e^{-\alpha}+e^{-\beta}\big)\partial_{i}\Psi\mathrm{d}u \mathrm{d}x^{i} + \delta_{ij} \mathrm{d}x^{i} \mathrm{d}x^{j}\,,
	\end{aligned}\label{}
\end{equation}
which remains flat.  The transformation also induces a background B-field, which from \eqref{OnnBginv} is
\begin{equation}
	\hB'_{0} = \frac{1}{4}\left(e^{-\alpha} - e^{-\beta}\right)\p_{i}\Psi \rd x^{i}\wedge\rd u 
	\equiv C'_{0}{}_{i}\rd x^{i}\wedge\rd u \, , \label{BtildeprimeCNM}
\end{equation}
where we have identified $C'_{0}$ from the Kaluza--Klein reduction of $\hB'_{0}$ \eqref{KKansatz_gBphi}.

We can also compute the transformed null vectors by first contracting \eqref{CNMlbarltu} with the inverse metric \eqref{gginvCNMut}, 
applying the transformation \eqref{double_boosts_jjbar} using \eqref{CNM_Oabcd} 
and then lowering the indices using the transformed background metric, as
\begin{equation}
	\hl'{}_{\hmu} = \hg'_{0}{}_{\hmu\hnu}\hl'{}^{\hnu} \, , \qquad \hbarl'{}_{\hmu} = \hg'_{0}{}_{\hmu\hnu}\hbarl'{}^{\hnu} \, .
\end{equation}
This gives the result
\begin{align}
	\hl' &= e^{-\alpha}\rd u \, , \\ \hbarl' &= e^{\beta}\rd t + \left[\frac{2F}{(F-1)}\tV_{i}  + \frac{1}{2}e^{\beta}\left(e^{-\alpha} + e^{-\beta}\right)\p_{i}\Psi\right]\rd x^{i} + \left[e^{-\alpha}Y - e^{-\beta}\frac{F^{2}}{(F-1)^{2}}\tV_{j}\tV^{j}\right]\rd u \, .
\end{align}

The transformation has a simple interpretation, as follows.  Define
\begin{equation}
	\Psi'(x^{j}) \equiv \frac{1}{2}\left(e^{-\alpha} + e^{-\beta}\right)\Psi(x^{j}) \, , \qquad Y' = e^{-\alpha-\beta}Y \, , \label{XYprimeCNM}
\end{equation}
and introduce the new coordinate
\begin{equation}
	\tv' \equiv t + \Psi'(x^{j}) + Y'u \, .
\end{equation}
The transformed metric and null vectors can be written succinctly in the $(u,\tv')$ coordinates as
\begin{equation}
\begin{aligned}
	&\rd\hs'_{0}{}^{2} = \rd u \rd\tv' + \delta_{ij}\rd x^{i}\rd x^{j} \, , \\
	 \hl' = e^{-\alpha}\rd u \, , \qquad \hbarl' &= e^{\beta}\rd\tv' + \frac{2F}{(F-1)}\tV_{i}\rd x^{i} + e^{-\beta}\frac{FK}{F-1}\rd u \, .
\end{aligned}
\end{equation}

From these ingredients we can reconstruct the total transformed metric,
\begin{equation}
	\hg'_{\hmu\hnu} = \htg'_{\hmu\hnu} - \frac{\kappa\varphi}{1 + \frac{1}{2}\kappa\varphi(\hl'\cdot\hbarl')}\cdot\frac{1}{2}\left(\hl'_{\hmu}\hbarl'_{\hnu} + \hl'_{\hnu}\hbarl'_{\hmu}\right) \, .
\end{equation}
Note that $\hl'\cdot\hbarl' = 2e^{\beta-\alpha} \neq \hl\cdot\hbarl$, so the scalar coefficient of the fluctuation part is not invariant under the transformation.  However, we may suggestively define
\begin{equation}
	\frac{F' - 1}{F'} \equiv e^{\beta - \alpha}\left(\frac{F - 1}{F}\right) \, , \qquad \tV'_{j} \equiv e^{-\alpha}\tV_{j} \, , \qquad
	K' \equiv -\frac{F'}{(F'-1)}\tV'_{j}\tV'{}^{j} = e^{-\alpha - \beta}K \, , \label{FVtildeKprime}
\end{equation}
from which we find
\begin{equation}
	\hbarl' = e^{\beta}\left[\rd\tv' + \frac{2F'}{(F'-1)}\tV'_{i}\rd x^{i} + \frac{F'K'}{(F'-1)} \rd u\right] \, , \qquad
	\frac{\kappa\varphi}{1 + \frac{1}{2}\kappa\varphi(\hl'\cdot\hbarl')} 
	= \left(1 - F'\right)e^{\alpha-\beta} \, .
\end{equation}
%
Further introducing a new coordinate by analogy with \eqref{vtilde},
\begin{equation}
	v' \equiv \tv' - \Psi'(x^{j}) = t + Y'u \, ,
\end{equation}
 we may write the transformed metric as simply
\begin{equation}
	\rd\hs'{}^{2} = F'\rd u\left[\rd v' + 2V'_{i}\rd x^{i} + K'\rd u\right] + \delta_{ij}\rd x^{i}\rd x^{j} \, , \label{CNM'metric}
\end{equation}
where we have defined, using \eqref{BtildeprimeCNM},
\begin{equation}
	V'_{i} \equiv e^{-\alpha}V_{i} - C'_{i} = \tV'_{i} + \frac{1}{2}\p_{i}\Psi' \, .
\end{equation}
The second equality may be verified explicitly using \eqref{BtildeprimeCNM} and \eqref{XYprimeCNM}, such that the definition is self-consistent.

Comparing with \eqref{CNMmetric}, we find that the transformation simply produces a different chiral null model metric.  Since $K'$, $V'_{i}$ and $F'{}^{-1}$ are linear functions of their respective unprimed versions, and by the homogeneity properties of the equations of motion \eqref{CNM_eom}, we may verify that this is indeed another chiral null model.  However, through the transformation we have induced a background B-field \eqref{BtildeprimeCNM} which is pure gauge.  Recall also that the KS decomposition is invariant under the field redefinition
\begin{equation}
	\hg'_{0} \rightarrow \hg'_{0} \, , \qquad \kappa\varphi \rightarrow e^{\beta - \alpha}\kappa\varphi \, , \qquad \hl' \rightarrow e^{\alpha}\hl' \, , \qquad \hbarl' \rightarrow e^{-\beta}\hbarl' \, . \label{CNMKSredefined}
\end{equation}
By applying this rescaling, we may recover the original form of our KS ansatz, now in terms of the transformed functions $F'$, $V'_{i}$ and $K'$.

Finally, from \eqref{CNM'metric} we may read off the transformed dilaton as
\begin{equation}
	\hphi' = \frac{1}{2}\ln F' = \frac{1}{2}\ln F + \frac{1}{2}(\alpha - \beta) - \frac{1}{2}\ln\left(1 + F(e^{\alpha-\beta} - 1)\right) \, ,
\end{equation}
where in the second equality we have used \eqref{FVtildeKprime}.  Note that for $\alpha = \beta$ the dilaton is invariant.

\subsubsection{Dimensional reduction}

Now let us turn to dimensional reduction of the chiral null model.  Our starting point is the KS decomposition in $(u,t)$ coordinates,
\begin{align}
	&\rd\hs_{0}{}^{2} = \rd u \rd t + \p_{i}\Psi\rd u \rd x^{i} + \rd u^{2} + \delta_{ij}\rd x^{i}\rd x^{j} \, , \qquad \hB_{0} = 0 \, , \qquad \kappa\varphi = F{}^{-1} - 1 \, , \label{CNMut_background} \\
	&\hl = \rd u \, , \qquad \hbarl = \left[\rd t + \left(\frac{2F}{(F-1)}\tV_{i} + \p_{i}\Psi\right)\rd x^{i} + \left(1 + \frac{F K}{(F-1)} \right) \rd u\right] \, , \label{CNMut_null}
\end{align}
where we have fixed $Y=1.$
Using \eqref{KKansatz_gBphi} the background metric reduces to
\begin{equation}
	\rd s_{0}{}^{2} = -\frac{1}{4}\left(\rd t + \p_{i}\Psi\rd x^{i}\right)^{2} + \delta_{ij}\rd x^{i}\rd x^{j} \, , \qquad  A_{0} = \frac{1}{2}\left(\rd t + \p_{i}\Psi\,\rd x^{i}\right) \, , \qquad h_{0}{}_{uu} = 1 \, . \label{CNMreducedmetric}
\end{equation}
For the fluctuation part, the null vectors decompose via \eqref{KK_null} into
\begin{equation}
\begin{aligned}
	&l = -\frac{1}{2}\left(\rd t + \p_{i}\Psi\rd x^{i}\right) 
	\, ,
	\\
	&\bar{l} = \frac{1}{2}\left[1 -\frac{FK}{F-1}\right]\left(\rd t + \p_{i}\Psi\rd x^{i}\right) + \frac{2F}{(F-1)}\tV_{i}\rd x^{i} \, ,
	\\
	&j_{u} = 1 \, , \qquad \bar{j}_{u} = 1 + \frac{FK}{F-1} \, .
\end{aligned} \label{CNMllbarjjbar}
\end{equation}

From these ingredients we can calculate the single copy of  the reduced chiral null model. It is obvious that the timelike Killing vector is covaraintly constant. After simplifying using \eqref{nulluindep} and \eqref{CNMreducedmetric}, we find the vector fields
\begin{equation}
	\mathcal{A} = \frac{1}{2}\left[K+\frac{1-F}{F}\right]l \, , \qquad \bar{\mathcal{A}} = \frac{(F-1)}{2F} \bar{l} = \mathcal{A} + \tV_{i}\rd x^{i} \, ,
\label{single_copy_CNM1}\end{equation}
and the internal scalars
\begin{equation}
	X_{u} = \frac{1}{2}\left[K + \frac{1 - F}{F}\right] \, , \qquad \bar{X}_{u} = \frac{1}{2}\left[K - \frac{1 - F}{F}\right] = X_{u} - \frac{1 - F}{F} \, .
\label{single_copy_CNM2}\end{equation}
We have seen in the previous subsection that the chiral null model is self-dual under the $\mathit{O}(1,1)^{2}$ double boost.  Thus applying this transformation amounts to replacing all variables with their primed versions.\footnote{Note that the induced background B-field \eqref{BtildeprimeCNM} is pure gauge and thus drops out of the equations of motion.}  


From \eqref{CNMreducedmetric} and \eqref{CNMllbarjjbar} we may reconstruct the total reduced fields using \eqref{KK_KS_undoubled}.  We find
\begin{equation}
\begin{aligned}
	\rd s^{2} &= -\frac{F}{4(K+1)}\left(\rd t + 2V_{i}\rd x^{i}\right)^{2} + \delta_{ij}\rd x^{i}\rd x^{j} \, , \qquad A = \frac{1}{2(K +1)}\left(\rd t + 2V_{i}\rd x^{i}\right) \, , \\ h_{uu} &= F(K + 1) \, , \qquad B = 0 \, , \qquad C = \frac{F}{2}\left(\rd t + 2V_{i}\rd x^{i}\right) \, , \qquad \phi = \frac{1}{4}\ln\left[\frac{F}{K + 1}\right] \, ,
\end{aligned} \label{CNMreduced_total}
\end{equation}
in agreement with \cite{Horowitz:1994rf}.\footnote{Interestingly, even though the KS ansatz becomes singular when $Y\rightarrow 0$, the total reduced fields are regular in this limit.  In \eqref{CNMreduced_total} this corresponds to the constant shift $K\rightarrow K-1$.}  Note that the pure-gauge contribution to $C$ has been removed by a gauge transformation: in the resulting gauge, \eqref{KKansatz_gBphi} implies that the four-dimensional B-field vanishes.

\subsubsection{Taub--NUT space in IWP class}
One example of the KK reduction of the CNM is a string theory generalisation of Taub--NUT space. If we choose the harmonic functions in the 5-dimensional CNM to be
\begin{equation}
  F^{-1} = 1+\frac{2m}{r}\,, \qquad K=1+\frac{2\tilde{m}}{r}\,, \qquad V_{\phi} = 2N\cos\theta \,,
\label{}\end{equation}
it corresponds to the 4-dimensional solution
\begin{equation}
  \mathrm{d} s^{2}=e^{4 \phi}(\mathrm{d}t+2 N \cos \theta \mathrm{d}\phi)^{2}-\left(d r^{2}+r^{2} \mathrm{d} \Omega^{2}\right) \, , \qquad e^{4 \phi}=\frac{1}{\left(1+\frac{2m}{r}\right)\left(1+\frac{2\tilde{m}}{r}\right)} \, ,
\label{}\end{equation}
where $N$ is the NUT charge.

In order to introduce a KS ansatz for this solution, we have to specify the $\tilde{V}_{i}$ by specifying $\Psi$. First, we assume that $\Psi$ is independent of the $\phi$ coordinate. We may solve the constraints and determine the $\tilde{V}_{i}$ and $\Psi$ function explicitly, giving
\begin{equation}
\begin{aligned}
  \tilde{V}_{r} &= -\frac{1}{2} \partial_{r} \Psi(r,\theta) = i \frac{\sqrt{4(N^{2}+m \tilde{m})+2mr}}{r}\,,
  \\
  \tilde{V}_{\theta} &= -\frac{1}{2} \partial_{\theta} \Psi(r,\theta) = -\frac{2N}{\sin\theta}\,,
  \\
  \tilde{V}_{\phi} &= V_{\phi} = 2i N \cos\theta \,,
\end{aligned}\label{}
\end{equation}
and
\begin{equation}
\begin{aligned}
  \Psi &= -4i \sqrt{4(N^{2}+m \tilde{m})+2mr} +8i \sqrt{m \tilde{m}+N^2} \coth^{-1}\left(\frac{2 \sqrt{m \tilde{m}+N^2}}{\sqrt{4(N^{2}+m \tilde{m})+2mr}}\right)
  \\
  &\quad +4 N \log \left(\tan\frac{\theta }{2}\right) \, .
\end{aligned}\label{}
\end{equation}
Note that there is no real solution for \eqref{nulluindep} and thus we have complexified $\tilde{V}_{i}$, which is common for Taub--NUT type solutions \cite{Luna:2015paa,Luna:2018dpt}. Before dimensional reduction, the KS ansatz is
\begin{equation}
\begin{aligned}
  \hat{l} &= \mathrm{d}u\,,
  \\
  \hat{\bar{l}} &= \mathrm{d}\tilde{v} - \Big(\frac{2\tilde{m}+r}{2m}\Big)\mathrm{d}u - \frac{i\sqrt{4(N^{2}+m \tilde{m})+2mr}}{m} \mathrm{d}r + \frac{2rN}{m \sin\theta} \mathrm{d}\theta -\frac{2iNr\cos\theta}{m} \mathrm{d}\phi\,,
  \\
  \kappa\varphi &= \frac{2m}{r}\,.
\end{aligned}
\end{equation}

Substituting the KK ansatz for $\hat{l}$ and $\hat{\bar{l}}$, we obtain
\begin{equation}
\begin{aligned}
  &l= -\frac{1}{2}\mathrm{d}t + i \frac{\sqrt{4(N^{2}+m \tilde{m})+2mr}}{r} \mathrm{d}r -\frac{2N}{\sin\theta} \mathrm{d}\theta\,, &\qquad j &= 1\,,
  \\
  &\bar{l} = \frac{2m+2\tilde{m}+r}{4m} \mathrm{d}t -\frac{2m+2\tilde{m}+3r}{4m}\big(\tilde{V}_{r}\mathrm{d}r +\tilde{V}_{\theta}\mathrm{d}\theta\big) - \frac{r}{m}\tilde{V}_{\phi}\mathrm{d}\phi\,, &\qquad \bar{j} &= \frac{2m-2\tilde{m}-r}{2m} \,.
\end{aligned}\label{}
\end{equation}
From \eqref{single_copy_CNM1} and \eqref{single_copy_CNM2}, the single copy for the solution is given, up to a $\mathit{U}(1)$ gauge transformation, by 
\begin{equation}
\begin{aligned}
  &\mathcal{A} = -\frac{m+\tilde{m}}{2r}\mathrm{d}t' \,,  &\qquad X_{u} &= \frac{m+\tilde{m}}{2r} \,,
  \\
  &\mathcal{\bar{A}} = -\frac{m+\tilde{m}}{2r}\mathrm{d}t' +2iN \cos\theta \mathrm{d}\phi\,, &\qquad \bar{X}_{u} &= \frac{-m+\tilde{m}}{2r} \,,
\end{aligned}\label{}
\end{equation}
where $t'$ is a new time coordinate defined by $\mathrm{d}t' = \mathrm{d}t+\frac{4N}{\sin\theta}\mathrm{d}\theta$.
Thus $\mathcal{A}$ and $\bar{\mathcal{A}}$ are the single copies for the Schwarzschild BH and the usual Taub-NUT space, respectively. 

\section{Conclusion} \label{sec:6}
This paper presents the Kaluza--Klein reduction of the Kerr--Schild ansatz for ungauged half-maximal supergravities and their T-duality transformations by using the relation with double field theory. Our starting point was $D$-dimensional NSNS supergravity without vector multiplets defined in a $T^{n}$-bundle with $n$ Abelian Killing vectors along the $T^{n}$ directions. Using the Kaluza--Klein ansatz for DFT, we constructed the Kerr--Schild ansatz for the lower-dimensional generalised metric as well as the half-maximal supergravity fields: metric, B-field, dilaton, $\mathit{U}(1)$ gauge fields and $\mathit{O}(n,n)$ moduli scalars. If we consider a flat background torus bundle, it is not appropriate to adopt a trivial background metric $\hat{\eta}$ when the background is nontrivial. Instead, we require a bundle patch given by a direct product between a base and a fibre patch in order to have a consistent KK reduction. However, the corresponding background metric is not constant in general. After KK reduction, the KS ansatz is nonlinear with respect to the fluctuation; however, the KS equations of motion are still linear.

Using the T-duality covariance of DFT, we constructed the $\mathit{O}(n,n)$ and $\mathit{O}(n+1,n+1)$ (when we include an isometry of the external space) transformations of the KS ansatz, including the background fields and the null vectors. Remarkably, the linear structure of the KS equations is preserved under T-duality. We considered the double Lorentz boosts $\mathit{SO}(1,1)^{2}$, a subgroup of $\mathit{O}(2,2)$ which preserves the asymptotic structure. We showed that the Buscher rule is realised by a simple sign-flipping transformation for a trivial background.

We showed that the single copy of a half-maximal supergravity is given by the KK reduction of the single copy of the higher-dimensional NSNS gravity, namely, a pair of higher-dimensional Yang--Mills theories. After KK reduction of $D$-dimensional NSNS supergravity, the single copy is the KK reduction of the abelian subsector of the corresponding super Yang-Mills theory. Remarkably, even though the lower-dimensional KS ansatz is nonlinear, the single copy is still proportional to the null vectors. Thus the T-duality transformation of the single copy is the same as that of the null vectors. If we consider a timelike Killing vector in the four-dimensional external spacetime, the electric and moduli scalar charges are mixed, but the magnetic charges are invariant under the double Lorentz boost transformation.

We applied our results to three examples: the Kerr black hole (BH), Sen's four-dimensional heterotic BH and the chiral null model. First, we obtained a new solution from the Kerr BH by applying the Buscher rule. Next, we constructed the corresponding KS ansatz and the classical double copy. Compared to the single copy of the Kerr BH, the only difference is the scalar and electric charges. Finally, we identified the single copy of the KK reduction of the chiral null model and showed that it is self-dual.

As is well-known, the KS ansatz cannot cover all solutions of half-maximal supergravity. There are some known extensions of the KS ansatz; however, it is not yet understood to what extent the exact double copy holds for generic spacetime. One natural extension of the KS double copy is to relax the null condition \cite{Ett:2010by,Kim:2019jwm}. However, the resulting equations of motion lose their linearity. That being said, we have shown that a particular combination of the equations of motion can linearise the field equations, which may provide a clue towards the general structure of the KS ansatz when the null condition is relaxed.

In this paper we have studied the KK reduction of ungauged supergravities only. It would be interesting to extend our results to half-maximal gauged supergravities using the generalised Scherk--Schwarz reduction of DFT. We may expect that the KS equations still remain linear even in the gauged case. This is particularly intriguing because it may shed insight into the structure of the non-abelian generalisation of the classical double copy. Furthermore, we should be able to find a KS-like ansatz for Yang--Mills theory by turning off the gravity multiplets. Another possible application is the generalisation to the half-maximal structure \cite{Malek:2017njj} in (heterotic) DFT, which yields $N=2$ supergravities in four dimensions.

\acknowledgments
We would like to thank Alejandro Rosabal for collaboration at the initial stage of this project. This work is supported by an appointment to the JRG Program at the APCTP through the Science and Technology Promotion Fund and Lottery Fund of the Korean Government. It is also supported by the Korean Local Governments of Gyeongsangbuk-do Province and Pohang City.


\newpage
\appendix

\section{General $\mathit{O}(D,D)$ transformation} \label{GeneralOnn}
A general $\mathit{O}(D,D)$ transformation $\hcO$ can be written as
\begin{equation}
\hcO_{\hM}{}^{\hN} = \begin{pmatrix}
a^{\hmu}{}_{\hnu} & b^{\hmu\hnu} \\
c_{\hmu\hnu} & d_{\hmu}{}^{\hnu}
\end{pmatrix} \, .
\end{equation}
This transformation should preserve the $\mathit{O}(D,D)$ invariant element $\hcJ$ as $\hcO\hcJ\hcO^{t} = \hcJ$, where
\begin{equation}
\hcJ = \begin{pmatrix}
0 & \delta^{\hmu}{}_{\hnu} \\
\delta_{\hmu}{}^{\hnu} & 0
\end{pmatrix} \, ,
\end{equation}
which implies the constraints
\begin{equation}
a b^{t}+ba^{t} = 0\,,\qquad a d^{t} + b c^{t} = \mathbf{1}_{D}\,, \qquad cd^{t} + dc^{t} = 0\,.
\label{Onnconstraints}
\end{equation}
By combining \eqref{Onnconstraints}, we can obtain other dependent expressions such as
\begin{equation}
a^tc + c^ta = 0 \, , \qquad b^td + d^tb = 0 \, , \qquad a^td + c^tb = \mathbf{1}_{D} \, . \label{Onnconstraintsalt}
\end{equation}

The DFT metric is written as
\begin{equation}
\hcH_{\hM\hN} = \begin{pmatrix}
\hg^{\hmu\hnu} & -\hg^{\hmu\hsigma}\hB_{\hsigma\hnu} \\
\hB_{\hmu\hrho}\hg^{\hrho\hnu} & \hg_{\hmu\hnu} - \hB_{\hmu\hrho}\hg^{\hrho\hsigma}\hB_{\hsigma\hnu}
\end{pmatrix} \, .
\end{equation}
$\hcH$ transforms linearly under $\mathit{O}(D,D)$ as
\begin{equation}
\hcH' = \hcO\hcH\hcO^t \, .
\end{equation}
Evaluating explicitly gives the nonlinear transformation rules for the component fields (generalized Buscher rules),
\begin{align}
\hg^{-1}{}' &= \left(a + b\hB\right)\hg^{-1}\left(a^t - \hB b^t\right) + b\hg b^t \, , \label{Buscherginv} \\
\hB'\hg^{-1}{}' &= \left(c + d\hB\right)\hg^{-1}\left(a^t - \hB b^t\right) + d\hg b^t \, , \label{BuscherBginv} \\
-\hg^{-1}{}'\hB' &= \left(a + b\hB\right)\hg^{-1}\left(c^t - \hB d^t\right) + b\hg d^t \, , \label{BuscherginvB} \\
\hg' - \hB'\hg^{-1}{}'\hB' &= \left(c + d\hB\right)\hg^{-1}\left(c^t - \hB d^t\right) + d\hg d^t \, . \label{BuschergBginvB}
\end{align}
Note that \eqref{BuscherBginv} and \eqref{BuscherginvB} are equivalent.  Define
\begin{equation}
\hE \equiv \hg + \hB \qquad \Rightarrow \qquad \hE^t = \hg - \hB \, . \label{hEEt}
\end{equation}
Using \eqref{Onnconstraints} and \eqref{hEEt} we may express \eqref{Buscherginv}, \eqref{BuscherBginv}, \eqref{BuscherginvB} and \eqref{BuschergBginvB} as
\begin{align}
\hg^{-1}{}' &= \left(a + b\hE\right)\hg^{-1}\left(a^t + \hE^tb^t\right)  \, , \label{ginvEform} \\
\hE'\hg^{-1}{}' &= \left(c + d\hE\right)\hg^{-1}\left(a^t + \hE^t b^t\right) \, , \label{EginvEform} \\
\hg^{-1}{}'\hE^t{}' &= \left(a + b\hE\right)\hg^{-1}\left(c^t + \hE^t d^t\right) \, , \label{ginvEtEform}\\
\hE'\hg^{-1}{}'\hE^t{}' &= \left(c + d\hE\right)\hg^{-1}\left(c^t + \hE^t d^t\right) \, . \label{EginvEtEform}
\end{align}
Inserting \eqref{ginvEform} into \eqref{EginvEform} and \eqref{ginvEtEform} gives
\begin{equation}
\hE' = \left(c + d\hE\right)\left(a + b\hE\right)^{-1} \, , \qquad \hE^t{}' = \left(a^t + \hE^tb^t\right)^{-1}\left(c^t + \hE^td^t\right) \, , \label{EEt}
\end{equation}
which is also consistent with \eqref{EginvEtEform}.

From \eqref{Onnconstraintsalt} we can derive another useful relation.  For any tensor $X_{\hmu\hnu}$,
\begin{align}
\left(c + dX\right)^t\left(a - bX^t\right) &= c^ta + X^td^ta - c^tbX^t - X^td^tbX^t \nn \\
&= -a^tc + X^t(\mathbf{1}_{D} - b^t c) - (\mathbf{1}_{D} - a^td)X^t + X^tb^tdX^t \nn \\
&= \left(a + bX\right)^t\left(-c + dX^t\right) \, , \label{transposeconstraint}
\end{align}
where in the second line we have used \eqref{Onnconstraintsalt}.   In particular, it follows that
\begin{equation}
\hE^t{}' = \left(a^t + \hE^tb^t\right)^{-1}\left(c^t + \hE^td^t\right) = \left(-c + d\hE^t\right)\left(a - b\hE^t\right)^{-1} \, . \label{Etalt}
\end{equation}
Defining
\begin{equation}
\cO_{1} \equiv a + b\hE \, , \qquad \cO_{2} \equiv a - b\hE^t \, , \qquad
\cO_{3}\hg \equiv c + d\hE \, , \qquad \cO_{4}\hg \equiv -c + dE^t \, , \label{O1234}
\end{equation}
we may write \eqref{EEt} and \eqref{Etalt} as
\begin{equation}
\hE' = \cO_{3}\hg\cO_{1}^{-1} \, , \qquad \hE^t{}' = \cO_{4}\hg\cO_{2}^{-1} \, . \label{hEEtOs}
\end{equation}

In terms of \eqref{O1234} the $\mathit{O}(D,D)$ constraints \eqref{Onnconstraints} become
\begin{align}
\cO_{1}\hg^{-1}\cO_{1}^t &= \cO_{2}\hg^{-1}\cO_{2}^t \, , \label{Onnab} \\
\cO_{3}\hg\cO_{3}^t &= \cO_{4}\hg\cO_{4}^t \, , \label {Onncd} \\
\mathbf{1}_{n} &= \frac{1}{2}\left[\cO_{1}\cO_{3}^t + \cO_{2}\cO_{4}^t\right] \, . \label{Onnadbc}
\end{align}
The constraint \eqref{transposeconstraint}, as applied to \eqref{Etalt}, can be written using \eqref{O1234} as
\begin{equation}
(\cO_{1}^t)^{-1}\hg\cO_{3}^t = \cO_{4}\hg\cO_{2}^{-1} \, . \label{O1324constraint}
\end{equation}
Combining \eqref{hEEtOs}, we see that the transformed metric and B-field are given by
\begin{align}
\hg' &= \frac{1}{2}\left[\cO_{3}\hg\cO_{1}^{-1} + \cO_{4}\hg\cO_{2}^{-1}\right] \, , \label{gprime} \\
\hB' &= \frac{1}{2}\left[\cO_{3}\hg\cO_{1}^{-1} - \cO_{4}\hg\cO_{2}^{-1}\right] \label{Bprime} \, .
\end{align}
From \eqref{O1324constraint} we may verify that $(\hg')^t = \hg'$ and $(\hB')^t = -\hB'$.

Consider transformations under the $\mathit{O}(1,D-1)^2$ subgroup of $\mathit{O}(D,D)$.  As discussed in section \ref{GlobalOnn}, we may parametrize such a transformation as
\begin{equation}
\hcO = \begin{pmatrix} a & b \\ c & d \end{pmatrix} = \frac{1}{2}\begin{pmatrix} o_{1} + o_{2} & \left(o_{1} - o_{2}\right)\eta^{-1} \\ \eta\left(o_{1} - o_{2}\right) & \eta\left(o_{1} + o_{2}\right)\eta^{-1} \end{pmatrix} \, , \label{Onn2xLorentzgeneral}
\end{equation}
where $\eta = \mathrm{diag}(-,+\ldots+)$.  In the case of a flat, Minkowski background, we have $\cO_{1}$ = $o_{1}$ and $\cO_{2} = o_{2}$, c.f. \eqref{jjbarflatprime}.  However, in the general case we find, explicitly,
\begin{align}
\cO_{1} &= \frac{1}{2}\left[o_{1}\left(\mathbf{1}_{D} + \eta^{-1}(\hg + \hB)\right) + o_{2}\left(\mathbf{1}_{D} - \eta^{-1}(\hg + \hB)\right)\right] \, , \label{cO1} \\
\cO_{2} &= \frac{1}{2}\left[o_{1}\left(\mathbf{1}_{D} - \eta^{-1}(\hg - \hB)\right) + o_{2}\left(\mathbf{1}_{D} + \eta^{-1}(\hg - \hB)\right)\right] \, , \label{cO2} \\
\cO_{3}\hg &= \frac{1}{2}\eta\left[o_{1}\left(\mathbf{1}_{D} + \eta^{-1}(\hg + \hB)\right) - o_{2}\left(\mathbf{1}_{D} - \eta^{-1}(\hg + \hB)\right)\right] \, , \label{cO3} \\
\cO_{4}\hg &= \frac{1}{2}\eta\left[-o_{1}\left(\mathbf{1}_{D} - \eta^{-1}(\hg - \hB)\right) + o_{2}\left(\mathbf{1}_{D} + \eta^{-1}(\hg - \hB)\right)\right] \, . \label{cO4}
\end{align}
From these we obtain
\begin{align}
\hg^{-1}{}' &= \frac{1}{4}\bigg[\left(o_{1} + o_{2}\right)\hg^{-1}\left(o_{1}^t + o_{2}^t\right) + \left(o_{1} - o_{2}\right)\eta^{-1}\hB\hg^{-1}\left(o_{1}^t + o_{2}^t\right) \nn \\ &\qquad\;\, - \left(o_{1} + o_{2}\right)\hg^{-1}\hB\eta^{-1}\left(o_{1}^t - o_{2}^t\right) + \left(o_{1} - o_{2}\right)\left(\hg - \hB\hg^{-1}\hg\right)\left(o_{1}^t - o_{2}^t\right)\bigg] \, , \label{ginvOform} \\
\hB'\hg^{-1}{}' &= \frac{1}{4}\eta\bigg[\left(o_{1} - o_{2}\right)\hg^{-1}\left(o_{1}^t + o_{2}^t\right) + \left(o_{1} + o_{2}\right)\eta^{-1}\hB\hg^{-1}\left(o_{1}^t + o_{2}^t\right) \nn \\ &\qquad\quad - \left(o_{1} - o_{2}\right)\hg^{-1}\hB\eta^{-1}\left(o_{1}^t - o_{2}^t\right) + \left(o_{1} + o_{2}\right)\left(\hg - \hB\hg^{-1}\hg\right)\left(o_{1}^t - o_{2}^t\right)\bigg] \, , \label{BginvOform}
\end{align}
\begin{align}
\hg' - \hB'\hg^{-1}{}'\hB' &= \frac{1}{4}\eta\bigg[\left(o_{1} - o_{2}\right)\hg^{-1}\left(o_{1}^t - o_{2}^t\right) + \left(o_{1} + o_{2}\right)\eta^{-1}\hB\hg^{-1}\left(o_{1}^t - o_{2}^t\right) \nn \\ &\qquad\quad - \left(o_{1} - o_{2}\right)\hg^{-1}\hB\eta^{-1}\left(o_{1}^t + o_{2}^t\right) + \left(o_{1} + o_{2}\right)\left(\hg - \hB\hg^{-1}\hg\right)\left(o_{1}^t + o_{2}^t\right)\bigg]\eta \, . \label{gBginvBOform}
\end{align}
Note that \eqref{ginvOform}, \eqref{BginvOform} and \eqref{gBginvBOform} are just the explicit forms of \eqref{Buscherginv}, \eqref{BuscherBginv} and \eqref{BuschergBginvB}, respectively, given in terms of \eqref{Onn2xLorentzgeneral}.  Combining the above, we may also write the transformations as
\begin{align}
\left(\mathbf{1}_{D} + \eta^{-1}\left(\hg' + \hB'\right)\right)\hg^{-1}{}'\left(\mathbf{1}_{D} + \left(\hg' - \hB'\right)\eta^{-1}\right) &= o_{1}\left(\mathbf{1}_{D} + \eta^{-1}\left(\hg + \hB\right)\right)\hg^{-1}\left(\mathbf{1}_{D} + \left(\hg - \hB\right)\eta^{-1}\right)o_{1}^t \, , \\
\left(\mathbf{1}_{D} - \eta^{-1}\left(\hg' + \hB'\right)\right)\hg^{-1}{}'\left(\mathbf{1}_{D} - \left(\hg' - \hB'\right)\eta^{-1}\right) &= o_{2}\left(\mathbf{1}_{D} - \eta^{-1}\left(\hg + \hB\right)\right)\hg^{-1}\left(\mathbf{1}_{D} - \left(\hg - \hB\right)\eta^{-1}\right)o_{2}^t \, , \\
\left(\mathbf{1}_{D} + \eta^{-1}\left(\hg' + \hB'\right)\right)\hg^{-1}{}'\left(\mathbf{1}_{D} - \left(\hg' - \hB'\right)\eta^{-1}\right) &= o_{1}\left(\mathbf{1}_{D} + \eta^{-1}\left(\hg + \hB\right)\right)\hg^{-1}\left(\mathbf{1}_{D} - \left(\hg - \hB\right)\eta^{-1}\right)o_{2}^t \, .
\end{align}
This identifies the field combinations which transform as tensors under $\mathit{O}(1,D-1)^2$.

Though we do not write explicit expressions for $\hg'$ and $\hB'$, these can be obtained straightforwardly for any given example as follows.  First compute the inverse metric using \eqref{ginvOform}.  Then take the inverse to obtain $\hg'$, and contract the resulting expression with \eqref{BginvOform} to find $\hB'$.  Alternatively, one may invert \eqref{cO1} and \eqref{cO2} in order to evaluate \eqref{gprime} and \eqref{Bprime}.

\newpage
\bibliography{references}
\bibliographystyle{JHEP}

\end{document}